\documentclass{aa}  

\usepackage{graphicx}
\usepackage{tikz}
\usepackage{siunitx}
\usepackage{txfonts}
\usepackage{amsmath}
\usepackage{amsfonts}
\usepackage{xcolor}
\usepackage{nicefrac}
\usepackage{microtype}
\usepackage{amssymb}
\usepackage{booktabs}      
\usepackage{tabularx}
\usepackage{pdflscape}
\usepackage{enumitem}
\usepackage[colorlinks = true,
           linkcolor = blue,
           urlcolor  = blue,
           citecolor = blue,
           anchorcolor = blue]{hyperref}
\usepackage[labelfont={bf,color=blue}]{caption}
\newcommand{\Gaia}{\textit{Gaia}\xspace}
\newcommand{\referee}[1]{#1}
\defcitealias{Hunt2024}{HR24}
\defcitealias{Childs2025}{Childs25}
\defcitealias{Almeida2023}{AMD23}

\begin{document} 

   \title{Stellar masses and mass ratios for \Gaia open cluster members}
   \titlerunning{Stellar masses and binary mass ratios for \Gaia DR3 open cluster members}

   \author{
    Sagar Malhotra
        \inst{\ref{inst:fqa},\ref{inst:iccub},\ref{inst:ieec}}\relax
    \and Alfred Castro-Ginard
        \inst{\ref{inst:fqa},\ref{inst:iccub},\ref{inst:ieec}}\relax
    \and Friedrich Anders
        \inst{\ref{inst:fqa},\ref{inst:iccub},\ref{inst:ieec}}\relax
    \and Carme Jordi           \inst{\ref{inst:ieec}}\relax
    \and Judit Donada
         \inst{\ref{inst:fqa},\ref{inst:iccub},\ref{inst:ieec}}\relax
    \and Xavier Luri
         \inst{\ref{inst:fqa},\ref{inst:iccub},\ref{inst:ieec}}\relax
    \and Lola Balaguer-Núñez
         \inst{\ref{inst:fqa},\ref{inst:iccub},\ref{inst:ieec}}\relax
    \and Songmei Qin
        \inst{\ref{inst:shanghai},\ref{inst:beijing},\ref{inst:iccub}}\relax
        \and Yueyue Jiang
        \inst{\ref{inst:shanghai},\ref{inst:beijing},\ref{inst:iccub}}\relax
    \and Andrija Župić
        \inst{\ref{inst:coruña},\ref{inst:iccub}}\relax
          }
    \institute{Departament de Física Quàntica i Astrofísica (FQA), Universitat de Barcelona (UB), C Martí i Franquès, 1, 08028 Barcelona, Spain\\
    \email{sagar@fqa.ub.edu}\relax
    \label{inst:fqa}
    \and{Institut de Ciències del Cosmos (ICCUB), Universitat de Barcelona (UB), C Martí i Franquès, 1, 08028 Barcelona, Spain \label{inst:iccub}}
    \and{Institut d'Estudis Espacials de Catalunya (IEEC), Edifici RDIT, Campus UPC, 08860 Castelldefels (Barcelona), Spain \label{inst:ieec}}
    \and{Key Laboratory for Research in Galaxies and Cosmology, Shanghai Astronomical Observatory, Chinese Academy of Sciences, 80 Nandan Road, Shanghai 200030, China \label{inst:shanghai}}
    \and{School of Astronomy and Space Science, University of Chinese Academy of Sciences, No. 19A, Yuquan Road, Beijing 100049, China \label{inst:beijing}}
    \and{Universidade da Coruña (UDC), Department of Computer Science and Information Technologies, Campus de Elviña s/n, 15071, A Coruña, Galiza, Spain \label{inst:coruña}}
    }

   \date{Received 2 October 2025; Accepted 5 December 2025}

  \abstract
  % context heading (optional)
   {Unresolved binaries in star clusters can bias stellar and cluster mass estimates, making their proper treatment essential for studying cluster dynamics and evolution.}
  % aims heading (mandatory)
   {We aim to develop a fast and robust framework for jointly deriving stellar masses and multiplicity statistics of member stars, together with optimal cluster parameters.}
  % methods heading (mandatory)
   {We use \Gaia DR3 parallaxes together with multi-band photometry of open cluster (OC) members to infer stellar masses and binary mass-ratios through simulation-based inference (SBI), while iteratively fitting the cluster parameters. The validation of our SBI framework on simulated clusters demonstrates that the inclusion of infrared photometry significantly improves the detection of low mass-ratio binaries. The minimum mass-ratio threshold for reliably identifying unresolved binaries depends on cluster properties and the available photometry, but typically lies below $q=0.5$.}
  % results heading (mandatory)
   {Applying our method to 42 well-populated OCs, we derive a catalogue of stellar masses and mass-ratios for 27\,201 stars, achieving typical uncertainties of 0.08 in $q$ and $0.01\,\mathrm{M}_\odot$ in the primary stellar mass. We analyse the archetype OCs M67 and NGC~2360 in detail, including mass segregation and mass-ratio distribution among other characteristics, while deriving multiplicity fractions for the rest of the sample. We find evidence that the high mass-ratio ($q\geq 0.6$) binary fraction shows a strong correlation with the age and a weak anti-correlation with the cluster metallicity. Furthermore, the variation of the binary fraction with stellar mass in OCs shows strong accordance with the observed dependence for field stars heavier than $\gtrsim0.6\,\mathrm{M}_\odot$.
   }
  % conclusions heading (optional)
   {Our work paves a path for future population-level investigations of multiplicity statistics and precision stellar masses in extended samples of OCs.}


   \maketitle
%
%-------------------------------------------------------------------

\section{Introduction}

Star clusters serve as natural laboratories for studying stellar evolution and dynamics, as they consist of coeval stellar populations with very similar chemical properties \citep[e.g.][]{Janes1982, Kalirai2010, Kamdar2019, Castro-Ginard2021}. They are also considered as the progenitors of most stars in the MW disc \citep{Lada2003, Krumholz2019}. On the other hand, multiplicity is widely recognized as a ubiquitous feature among the Galactic field stellar population \citep{Offner2022} with about 50\% solar-type stars observed as part of multiple systems \citep[e.g.][]{Fuhrmann2017, Moe2017}. Studying multiplicity in open clusters (OCs) can therefore provide valuable insights into the origin and evolution of binary star systems \citep[e.g.][]{Ebrahimi2022, Cordoni2023}. Moreover, diverse stellar environments in clusters can help us understand how binary system statistics such as multiplicity fraction and mass-ratio (\textit{q}) distribution evolve \citep{Cournoyer-Cloutier2021, Guszejnov2023, Pavlik2025}, in addition to providing crucial insights about linking the formation mechanisms of such systems with their observed present-day properties in the Galactic field \citep{Goodwin2005, Parker2009}.

Historically, unresolved binaries have been linked to increased scatter along the main-sequence (MS) in colour-magnitude diagrams (CMDs) \citep{Schlesinger1975}, with \citet{vandeKamp1968} and \citet{Bettis1977} being among the first to discuss their prevalence in OCs. Classical studies such as \citet{Duquennoy1991, Raghavan2010, Moe2017} provided baseline statistics for binary fractions in the field. In parallel, efforts using photometric methods in globular clusters \citep{Ivanova2005, Milone2012} and N-body simulations \citep{Heggie1975, Hurley2007, PortegiesZwart2001, Kaczmarek2011} have shown that binary fractions evolve dynamically over time due to mass segregation, tidal stripping, and stellar exchange encounters. These effects can alter the binary population along with the present-day mass function, making simple corrections for unresolved systems often insufficient when estimating OC mass distributions and total masses \citep{Rastello2020}.

With its unprecedented precision, the {\it Gaia} mission \citep{GaiaCollaboration2016} has revolutionized our ability to detect multiplicity signatures through photometry, astrometry, and spectroscopy \citep{GaiaCollaboration2018Babusiaux, Belokurov2020, GaiaCollaboration2023Arenou, Castro-Ginard2024, Li2025}. This opens new avenues to study unresolved binaries in OCs across the Milky Way, where detection methods previously relied on time-consuming spectroscopic \citep{Torres2020, Torres2021} or eclipsing binary surveys \citep{Mermilliod1992, Mazur1995, Brewer2016, Brogaard2021}.

The impact of binaries on the dynamics of OCs has been studied using N-body simulations intensively in the past 50 years \citep{Heggie1975, Hurley2005, Marks2011, Geller2013, Geller2015}. For example, \citet{Parker2009} predicted that wide binaries should be destroyed after a few crossing times, a fact that was later corroborated with {\it Hubble} data for globular clusters \citep{Milone2012} and \Gaia DR2 data for OCs \citep{Deacon2020}.
The advent of \Gaia has already allowed for several in-depth studies of binaries in individual OCs using multi-band photometry \citep[e.g.][]{Cohen2020, Li2020, Malofeeva2022, Malofeeva2023, Childs2024}. Larger samples ($\sim 200$) of OCs with binary statistics are also starting to be assembled \citep[e.g.][]{Donada2023, Jiang2024}. Mostly, however, the modelling does not allow for precise determinations of individual mass-ratios. In this paper, we attempt to obtain precise individual stellar masses and mass-ratios of unresolved binaries in OCs using a self-consistent (although still unavoidably model-dependent) way. 

Our paper is structured as follows: In Sect. \ref{sec:data} we summarise the data along with the selection of the sample of OCs used in this study. Section \ref{sec:pipleline} describes the methodology we employ to simultaneously infer individual stellar masses, mass-ratios of unresolved binaries, as well as cluster parameters such as age, distance, and interstellar extinction. In Sect. \ref{sec:sim_clus_performance} we test our methodology on simulated star clusters, proceeding to present the results for the selected {\it Gaia} DR3 OCs in Sect. \ref{sec:results} (readers mostly interested in the science results might want to jump directly to Sect. \ref{sec:results}). We provide a  comparison with the literature in Sect. \ref{sec:comp_with_lit} and discuss the results in a broader context in Sect. \ref{sec:discussion}. Our conclusions are presented in Sect. \ref{sec:conclusions}.

%--------------------------------------------------------------------
\section{Data}\label{sec:data}

In this paper, we make use of multi-band photometry\footnote{The multi-band photometric observations are compiled by crossmatching \Gaia sources with the 2MASS and WISE catalogues, the details of which are presented in App.~\ref{app:2mass_wise_crossmatch}.} from the Two Micron All Sky Survey (2MASS) \citep{Skrutskie2006}, Wide-field Infrared Survey Explorer (WISE) \citep{Wright2010} and \Gaia along with \Gaia DR3 parallax measurements of stars in OCs. To maximize homogeneity, we use the OC membership lists provided by \citet[][hereafter HR24]{Hunt2024} for 7167 cluster/moving group candidates of which 3530 are classified as reliable clusters\footnote{CST > 5.0; \texttt{class\_50}  $>$ 0.5; \texttt{kind} $=$ "o"} by the authors. We also include 2 objects UPK~303 and UPK~612 (classified as moving groups by \citetalias{Hunt2024}) owing to their very well-defined CMDs. Further, only the OCs\footnote{Even though our sample contains 2 moving groups classified by \citetalias{Hunt2024}, we use the term "OCs" in a broader sense.} within 2~kpc from the Sun are considered in order to limit the impact of parallax uncertainties and interstellar extinction.

Although several previous studies have estimated the ages, distances, and line-of-sight extinctions of clusters \citep[e.g.][]{Bossini2019, Cantat-Gaudin2020, Dias2021, Hunt2023, Cavallo2024}, we find that many of these solutions do not optimally reproduce the observed CMDs, rendering them unsuitable for a robust characterization of unresolved binaries. Hence, we re-estimate the cluster parameters (see Sect.~\ref{subsec:iter_cluster_param}) while using spectroscopically derived metallicities compiled by \citet{Joshi2024}. In addition, we select OCs with well-defined, densely populated main sequences and no evidence of differential extinction in their CMDs, yielding 144 candidates. We classify these OCs into three categories--`Poor,' `Good,' and `Very Good'--based on the reliability of the cluster parameters inferred by our method. Of the 144 candidates, 42 ($\sim32\%$) fall into the `Very Good' category and are the focus of this study (see Sect. \ref{sec:results}).

%--------------------------------------------------------------------

\section{Method}\label{sec:pipleline}

\begin{figure*}
   \centering
            \includegraphics[width = 1.0\textwidth]{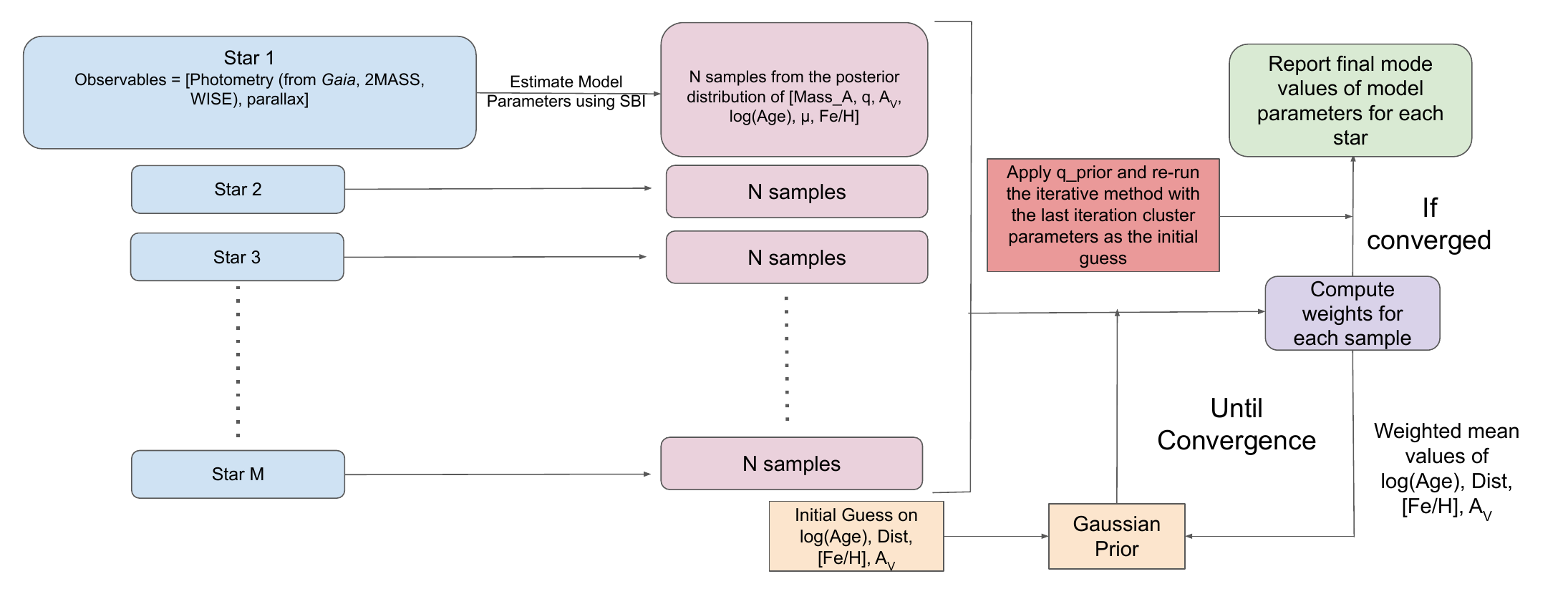}
   \caption{A schematic diagram showing the workflow of obtaining the joint-posterior distribution of stellar mass and mass-ratio while iteratively updating the cluster parameters until convergence. We infer posteriors using SBI for all open cluster members (Sect.~\ref{subsec:model_param_sbi}) and iteratively improve the estimated cluster parameters, log(Age), distance and $A_V$ (Sects.~\ref{subsec:apply_cluster_priors} and ~\ref{subsec:iter_cluster_param}). We also make use of the empirical findings of binary fraction varying with stellar mass and incorporate it in a mass-ratio prior (Sect.~\ref{subsec:mass_ratio_prior}).}
   \label{fig:iterative_schematic}
\end{figure*}

Our pipeline consists of the following: (i) employing Simulation-Based Inference \citep[SBI, e.g.][]{Greenberg2019, Papamakarios2019, Cranmer2020} to infer stellar parameters (Sect.~\ref{subsec:model_param_sbi} and \ref{subsec:simulate_training_data}), (ii) estimating cluster parameters through an iterative procedure (Sect.~\ref{subsec:apply_cluster_priors} and \ref{subsec:iter_cluster_param}), and (iii) using the observed mass dependence of binary fraction (Sect.~\ref{subsec:mass_ratio_prior}) to refine the joint mass–$q$ posterior. We show a schematic diagram for the adopted methodology in Fig.~\ref{fig:iterative_schematic}.

\subsection{Simulation-Based Inference (SBI) of stellar parameters}\label{subsec:model_param_sbi}

We aim to infer the posterior distribution of astrophysical parameters for each star from its observed multi-band photometry provided by \Gaia, 2MASS and WISE, in addition to the \Gaia parallax ($\varpi$). The parameters include primary mass ($M_{\mathrm{A}}$), metallicity ([$\mathrm{Fe/H}$])\footnote{We assume $[\mathrm{M/H}] = [\mathrm{Fe/H}]$ throughout this study, which is reasonable for OCs in the solar vicinity.}, age (log$_{10}$(Age [yr])), mass-ratio ($q$), distance modulus ($\mu$), and line-of-sight extinction at 550~nm ($A_V$). We adopt a similar method as proposed by \citet{Wallace2024} and employ SBI, which avoids the need for an explicit likelihood function. In particular, we apply Neural Posterior Estimation \citep[NPE; e.g.][]{Papamakarios2016}, in which a neural network is trained on simulated data to parametrize a density estimator that approximates the posterior distribution. This approach allows for a significant speedup over traditional inference methods and can be efficiently scaled over a large sample, making it particularly useful for homogeneous studies on OC members.

\subsection{Simulated training dataset}\label{subsec:simulate_training_data}

In order to test the influence of systematic uncertainties in stellar evolutionary models, we test two sets of stellar models, namely the PARSEC \citep{Bressan2012} and the MIST \citep{Dotter2016, Choi2016} libraries. However, owing to notable differences in the predicted photometry of low-mass stars between the two sets of isochrones \citep[e.g.][]{Choi2016}, we adopt the PARSEC isochrones in this work to facilitate a direct comparison with previous studies, although our approach is readily applicable with any theoretical stellar model. We simulate the corresponding observables of about 2 million stars, \referee{broadly} resembling a typical Galactic \referee{cluster} population, with the prior distributions taken from Table~\ref{table:priordist}. \referee{We adopt these broad priors to include almost all physically plausible combinations of the respective parameters so that our method can be applied to previously unseen data without introducing significant systematic biases.} The apparent magnitudes are obtained by assigning a distance, $d$~ (pc), and a line of sight extinction, $A_V$~ (mag), while using the extinction coefficients\footnote{\href{https://www.cosmos.esa.int/web/gaia/edr3-extinction-law}{https://www.cosmos.esa.int/web/gaia/edr3-extinction-law}} for \Gaia, 2MASS \citep[following][]{Danielski2018} and WISE passbands \citep{Yuan2013}. The true parallax, $\varpi$, is simply computed as the inverse of the simulated distances. While simulating the training dataset, we occasionally generate non-physical stars (e.g. stars too massive for the simulated age) or very faint stars (i.e. $G\gtrsim21$~mag) which are mostly not detected by \Gaia. Hence, these stars are subsequently removed from the training data. Moreover, as shown in Table~\ref{table:priordist}, we use a lower limit of 0.1~$\mathrm{M}_{\odot}$ on the stellar mass for our simulations. Therefore, in cases where the companion star’s mass is lower than this threshold, the system is treated as a single star, resulting in an excess of simulated systems at mass ratio $q = 0$. Figure~\ref{fig:prior_distribution} shows the resulting distribution of astrophysical parameters of simulated unresolved binaries and single stars.

Since SBI requires the training data to closely resemble the real data \citep{Kelly2025}, we incorporate observational uncertainties by directly injecting noise into the simulated observables, rather than treating the uncertainties as separate variables during training. This approach helps us avoid poorly assuming that the mean values of simulated observables are representative of the true ones. We use a similar method described in \citet{Wallace2024} to simulate uncertainties in the photometric magnitudes and parallaxes by simply interpolating their observed dependence on the apparent magnitudes in the real data \citep[e.g.][]{Fabricius2021}. 

We note that there have been numerous studies about the colour deviation between the observed \Gaia CMDs of OCs and the theoretical isochrones \citep[e.g.][]{Wang2025}, especially in the low-mass regime ($M \sim 0.3\,-\,0.6\,\mathrm{M}_{\odot}$). This might affect our results for a few OCs in the mass range of our interest, and we discuss this in Sects.~\ref{sec:results} and \ref{sec:discussion}.

\begin{table}
\caption{Prior distributions for model parameters used for generating the simulated training dataset.}  
\label{table:priordist}
\centering              
\begin{tabular}{c c c}       
\hline\hline        
Param. & Range & Distribution \\
\hline                      
   $M_{A}$ & [0.1,\, 9.0]& $\mathcal{B}(1.0,\, 5.0)$ \\ 
   $[\mathrm{Fe/H}]$ & [-2.0,\, 0.45] &   $\mathcal{B}(10.0,\, 2.0) - 0.9^{(a)}$   \\
   log(Age) & [6.5,\, 10.0]&    $\mathcal{U}(7.0,\, 10.0)$   \\
   $q$ & [0.0, \, 1.0] &   $\mathcal{U}(0.0,\, 1.0)$   \\
   $\mu$ & [0.0, \, 13.5]& $\mathcal{U}(0.0,\, 13.5)$\\
   $A_{V}$ & [0.0, \, 4.0]& $\mathcal{U}(0.0,\, 4.0)$\\
\hline                                 
\end{tabular}
\tablefoot{$\mathcal{U}(a,\, b)$ and $\mathcal{B}(a,\,b)$ indicate the uniform and the beta distributions, respectively. (a) 0.9 is subtracted so that the metallicity distribution of our training dataset peaks close to the solar metallicity.}
\end{table}

These simulations are then used to train the neural network and obtain the posterior distributions using SBI\footnote{\href{https://sbi-dev.github.io/sbi/0.22/}{https://sbi-dev.github.io/sbi/0.22/}}, provided the photometric and parallax measurements of \citetalias{Hunt2024} cluster members. However, as multi-band photometry from the three surveys is not available for each cluster member, we derive three posteriors, each corresponding to different subsets based on the availability of photometric measurements (see Fig.~\ref{fig:g_hist_surveys}). After multiple attempts, we found that 20\,000 posterior samples\footnote{If posterior sampling with the full set of available observables fails, we use progressively reduced number of passbands, while employing only \Gaia photometry on the last attempt.} for each star work best in the current scenario.

\begin{figure}
    \centering    
    \includegraphics[width=0.48\textwidth]{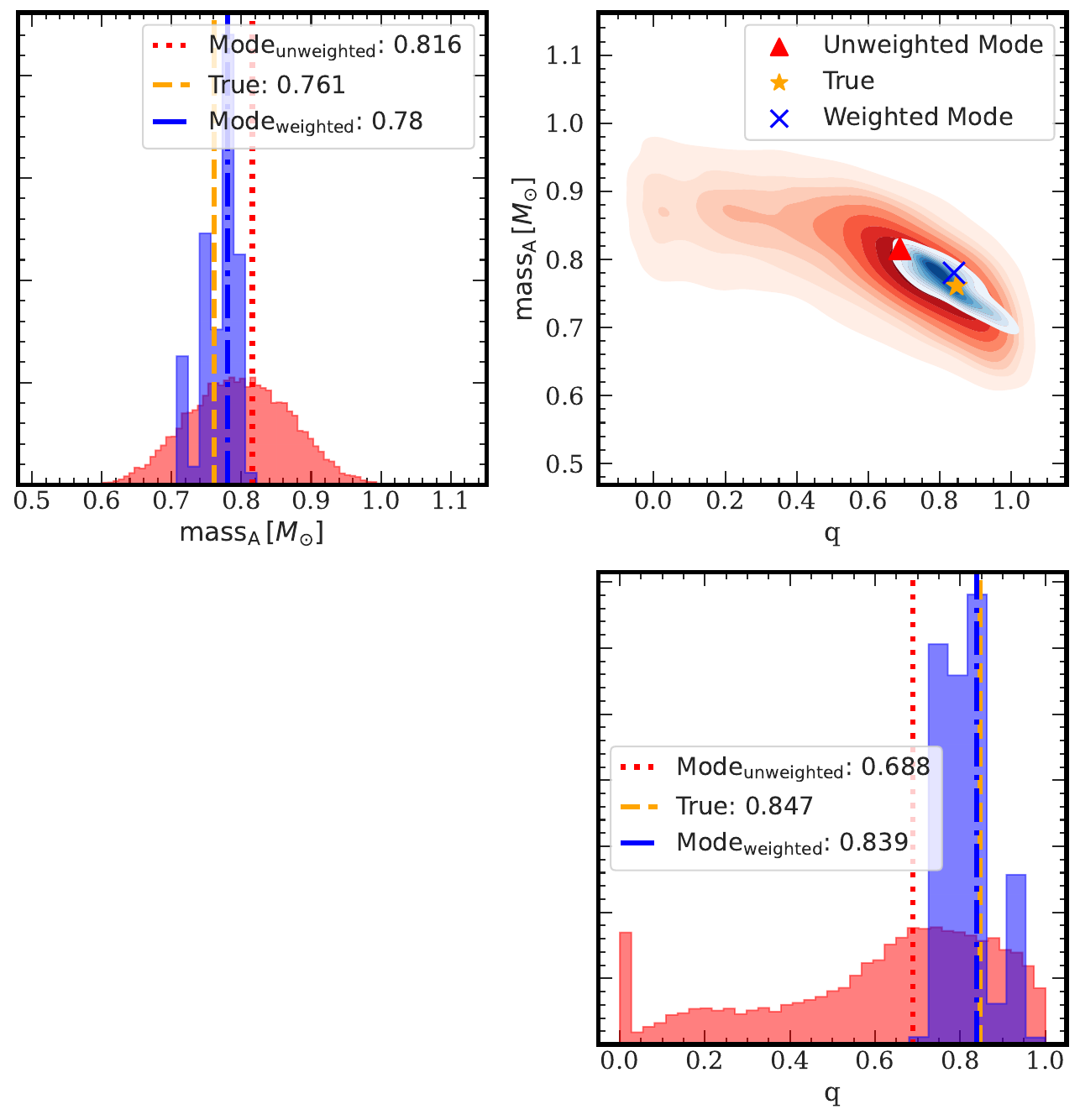}
    \caption{Corner plots of SBI posteriors for a simulated low-mass binary star ($M_{\mathrm{A}}\sim0.76\,\mathrm{M}_{\odot}, q\sim0.85, d=1$~kpc, $A_V\sim0.48$ mag, log(Age)~[dex] $=9.5$, [Fe/H] [dex] = 0.0) with typical \Gaia DR3 and 2MASS/WISE uncertainties for two cases: without (red) and with (blue) prior information about cluster parameters to obtain the weighted posterior distribution.}
\label{fig:sim_star_corner_plot}
\end{figure}

\subsection{Applying cluster parameter priors}\label{subsec:apply_cluster_priors}

As shown in Fig.~\ref{fig:iterative_schematic}, we incorporate the cluster parameter information by assigning weights to SBI posterior samples of cluster members while assuming a Gaussian distribution in log(Age), distance, [Fe/H] and $A_V$. These Gaussian distributions are truncated at $\pm\,5\sigma$ leading to some photometric outliers that have posteriors inconsistent with the assumed cluster parameters. In this entire procedure, the resulting weight assigned to each posterior sample is approximated by a multiplication of the individual weights corresponding to each Gaussian prior. Though this is not perfectly ideal due to expected correlations among the derived cluster parameters, with the correct way being sampling from the conditional posterior while cluster parameters are kept constant, we do not adopt this due to the following: a)~We aim to simultaneously infer the log(Age), distance and $A_V$ without using any prior information; b)~Limited availability of computational resources required for the traditional Markov-Chain Monte-Carlo (MCMC) method used in such a sampling; c)~As shown in App.~\ref{app:conditional_sbi}, our method of assigning weights to posterior samples does not significantly affect the results.

Figure~\ref{fig:sim_star_corner_plot} illustrates the use of prior information about the cluster parameters in the inference of masses and mass-ratios of cluster members. The red contours show the (unweighted) posterior distributions of the astrophysical parameters obtained using SBI while the blue ones depict the weighted distribution of the posterior samples after applying cluster parameter priors. The resulting modes of the weighted posterior distributions recover the underlying simulated masses and mass-ratios with high precision, in contrast to the case where the cluster parameter information is not incorporated. It is worth noting that in the case of weighted posteriors, out of a total of 20\,000 samples, only 175 samples had a non-zero weight, which is slightly below the typical valid sample rate of about 4\%.

\subsection{Mass-ratio prior}\label{subsec:mass_ratio_prior}

After applying the cluster parameter priors, we also make use of the empirical findings that the binary fraction increases monotonically with the primary stellar mass (\citealt{Moe2017}, \citealt{Offner2022}). We introduce a suitable prior on the joint $M_{\mathrm{A}}$-$q$ posterior distribution by assigning weights to different regions of the mass-ratio posterior based on the empirical probability of a star to be in a binary system. In other words, $q$ posterior samples indicative of binary systems are given higher weights if the corresponding primary star is more likely to have a companion and lower weights otherwise. For this, we need two ingredients:~a) an analytical fit to the observed dependence of binary fraction on the stellar mass, b) a lower limit on a mass-ratio below which all posterior samples are considered to correspond to a single star and vice versa.

We use the following equation provided by \citet{Arenou2011} that roughly fits the observed variation of the binary fraction w.r.t the $M_\mathrm{A}$:
\begin{equation}
    f^{\mathrm{emp}}_{b}(M_\mathrm{A})=0.8388\cdot\tanh(0.688\,M_{\mathrm{A}} + 0.079)
\label{eq:binary_fraction_empirical}
\end{equation}
where $f^{\mathrm{emp}}_{b}$ denotes the observed binary fraction of stars with a particular mass $M_{\mathrm{A}}$ in the Galactic field. 
Secondly, the limiting value of the mass-ratio, $q_{\mathrm{lim}}$, can be estimated using the minimum value of stellar mass used in the simulations i.e. 0.1~$\mathrm{M}_{\odot}$ (Sect.~\ref{subsec:simulate_training_data}). Since any binary companion with a mass lower than this value will be considered as a "dark component" by the SBI, this sets a value of $q_{\mathrm{lim}}$ for each mass: $q_{\mathrm{lim}} = \frac{0.1}{M_\mathrm{A}}$, that is not only intrinsic to our simulations but also holds physical meaning (in which case it corresponds to $\frac{\sim0.075}{M_\mathrm{A}}$ where 0.075~$\mathrm{M}_{\odot}$ represents the usual stellar mass limit). This method of using a $q_{\mathrm{lim}}$ based on stellar models/simulations has also been used before by \citet{Rongrong2025}. 

Thus, using the two ingredients, we can simply construct the $q_{\mathrm{prior}}$ as:

\begin{equation}
    q_{\mathrm{prior}} = 
    \begin{cases}
        \frac{1 - f^{\mathrm{emp}}_{b}}{q_{\mathrm{lim}}}, &  \hspace{0.0cm} q < q_{\mathrm{lim}}, \\
        \\
        \frac{f^{\mathrm{emp}}_{b}}{1 - q_{\mathrm{lim}}}, &  
        \hspace{0.0cm} q \geq q_{\mathrm{lim}},
    \end{cases}
    \label{eq:q_prior}
\end{equation}

As illustrated in Sect.~\ref{subsec:apply_cluster_priors} and Fig.~\ref{fig:sim_star_corner_plot}, we adopt the mode of the weighted posterior distribution as the estimate of recovered model parameters. Unlike other summary statistics such as the mean or median, the mode can be ambiguous, with multiple, infinitely many, or even undefined values. To obtain a well-defined estimate of the mass-ratio for binary classification, we use the fraction of weighted posterior samples corresponding to the binary star case as a guess for computing the mode. This fraction can be interpreted as the probability, $P_{b}$, that a given star is an unresolved binary. We set $q_{\mathrm{mode}}$ to a value greater than $q_{\mathrm{lim}}$ if $P_{b} > 0.5$ (i.e. the star is more likely to be an unresolved binary), and to $0.0$ otherwise. For remaining parameters, we adopt the "global" mode, defined as the highest weighted bin by applying the Sturges \citep{Sturges01031926} or Freedman-Diaconis \citep{Freedman1981OnTH} binning rule to their respective posterior distributions.

\subsection{Estimating cluster parameters: An iterative method\label{subsec:iter_cluster_param}}

\begin{figure}
    \centering    
    \includegraphics[width=0.5\textwidth]{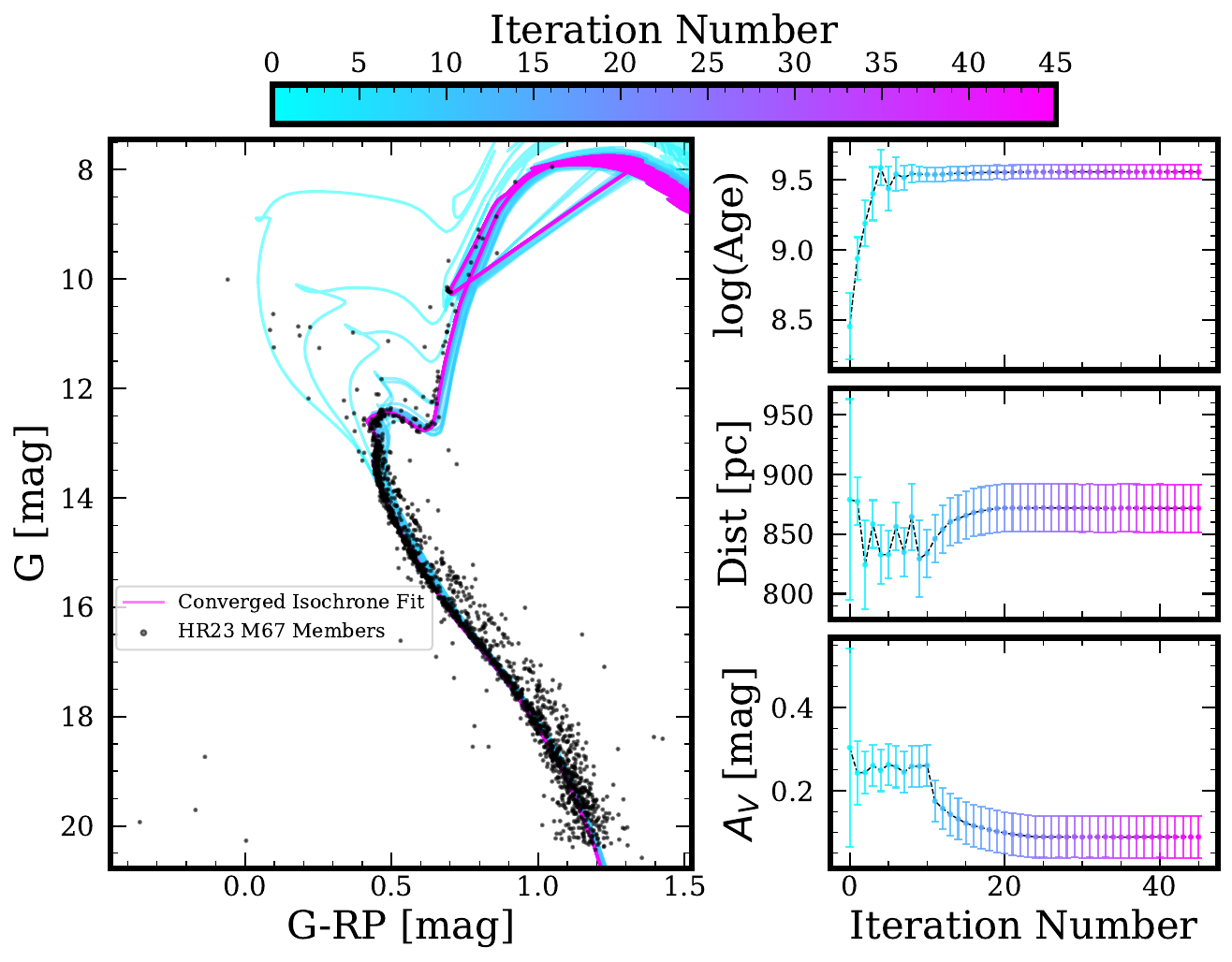}
    \caption{Visualisation of the convergence of the iterative fitting loop (see Fig.~\ref{fig:iterative_schematic}) for the case of M67 (NGC 2682). The left panel shows the PARSEC isochrone fits to the observed CMD for each iteration while the panels on the right illustrate the variation of cluster parameters across iterations before they converge in the first round at the 45th iteration.}
\label{fig:cluster_param_iter_method}
\end{figure}

Our approach of deriving optimal cluster parameters is based on the following: iterative selection of subsets of stars in the cluster CMD, that are best suited for estimating specific cluster parameters, while refining the weights of their posteriors at each step until convergence is achieved. As a result of the challenges in accurately determining metallicity from photometry and in order to limit degeneracy among the inferred parameters, we use the spectroscopic metallicity estimates of OCs (see Sect.~\ref{sec:data}) while keeping it fixed throughout the iterative fitting loop. For brevity, we detail the entire procedure in App.~\ref{app:cluster_param_iter_method} and only show how we are able to recover the optimal cluster parameters for the well-known cluster M67 in Fig.~\ref{fig:cluster_param_iter_method}. We iterate through the first loop until convergence\footnote{both parameters and the corresponding uncertainties do not change by more than 0.1\% throughout the last 5 iterations.} where the initial guess is taken as the mean of relevant unweighted SBI posteriors of all the stars in M67 membership list. As evident in Fig.~\ref{fig:cluster_param_iter_method}, the initial guess significantly underestimates the age due to the presence of blue stragglers. However, our automated selection of stars based on their evolutionary phase selects only the main-sequence turnoff (MSTO) stars while specifically removing the blue stragglers which results in a significant improvement of the estimated cluster age within the first few iterations. Other parameters such as the distance and $A_V$ are also improved until they all converge after 45 iterations. 

The convergence of the estimated parameters in the first loop is followed by a subsequent run of the method, using the values of the final iteration as the initial guess and incorporating the application of $q_{\mathrm{prior}}$.

 %--------------------------------------------------------------------

\section{Performance for simulated clusters}\label{sec:sim_clus_performance}

\subsection{Cluster simulations}

We test the performance of our methodology on simulated star clusters with controlled binary properties and typical \Gaia DR3-like uncertainties. Using PARSEC isochrones 1.2S, we simulate coeval, chemically homogeneous stellar populations of 500 cluster members with primary masses randomly sampled from a Kroupa IMF \citep{Kroupa2001} (with a lower mass bound of 0.1~$\mathrm{M}_{\odot}$). The probability of a star being in a binary system was governed by Eq.~\ref{eq:binary_fraction_empirical} and the corresponding secondary masses of those systems were simulated by randomly sampling mass-ratios from a uniform distribution between $\frac{0.1}{M_{\mathrm{A}}}$ and $1.0$. The combined flux of each binary system was used to compute the absolute magnitude before placing the simulated star cluster at a particular distance with an associated line-of-sight extinction to obtain the desired observables for each star. It should be emphasized that even though we simulated typical \Gaia DR3-like uncertainties in the observables, these simulated clusters will still represent an idealized scenario since we did not model stellar rotation and differential extinction, both of which affect the position of a star in the CMD \citep{Platais2012}. Moreover, we also did not try to model potential systematic errors in the synthetic photometry of PARSEC models (e.g. \citealt{Brandner2023}).

\subsection{Joint ${M_{\mathrm{A}}}$-q posteriors for simulated clusters}

\begin{figure*}
\sidecaption{}
\centering
\includegraphics[width=0.7\textwidth]{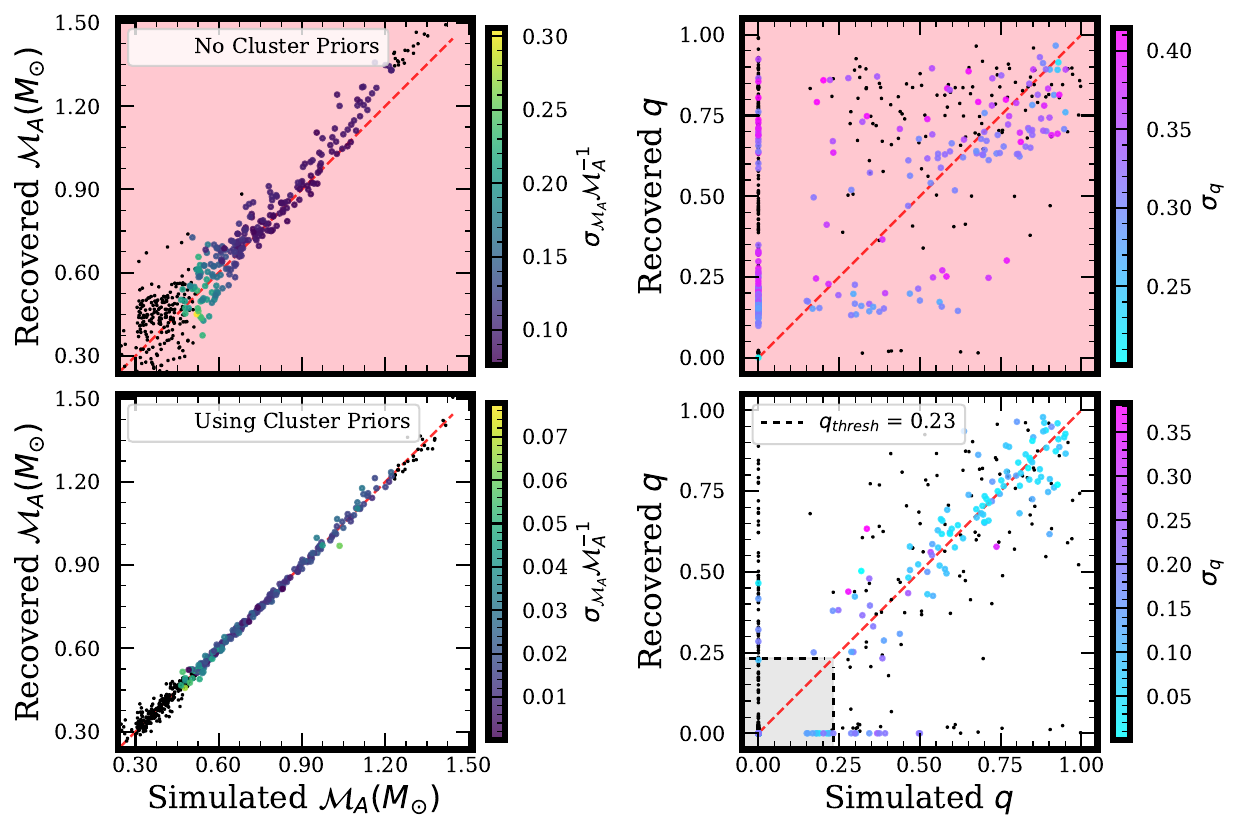}
\caption{Recovered masses and mass-ratios of stars in a simulated solar-metallicity cluster ($d=1$~kpc, $A_V\sim0.1$ mag, log(Age) $=9.5$ dex). Modes of the unweighted (\textit{Top}) and weighted (\textit{Bottom}), using cluster parameter information, $M_\mathrm{A}$ and $q$ posteriors are shown w.r.t the simulated parameters. The colour of each data point represents the (fractional) uncertainty in the corresponding ($M_\mathrm{A}$) $q$, whereas black data points correspond to stars that are not in the MS or have $G\geq19$ mag. The dashed line in the bottom right plot indicates the $q_{\mathrm{thresh}}$ (see text for details), above which we can reliably predict the mass-ratio of an unresolved binary in this cluster.}
\label{fig:clus_performance_sbi}
\end{figure*}

As an example, we show in Fig.~\ref{fig:clus_performance_sbi} the recovered values of the masses and mass-ratios of individual stars in a simulated cluster, while using multi-band photometry along with the \Gaia parallax, for two cases: a) before (top, shaded) and b) after (bottom) applying Gaussian priors in cluster parameters. It is evident that the one-to-one comparison plots are significantly improved when we use cluster parameter information to obtain the weighted $M_{\mathrm{A}}$ and $q$ posteriors while also improving the precision, depicted by the colour of each point, by almost an order of magnitude. Hence, incorporating such information about the host cluster not only helps in reducing the degeneracies, but also constrains the location of the single star MS on the CMD. The top panel of this figure indicates that treating stars as part of the field leads to both the masses and mass-ratios to be significantly uncertain. In contrast, the bottom panel shows that the masses of stars are reliably predicted across a wide range of evolutionary phases by the mode of the posterior distribution weighted by cluster parameter priors. However, we cannot satisfactorily recover the mass-ratios for stars not in the MS or fainter than $G=19$ mag (indicated by small black points in the plot), due to the very small light ratio in the case of a MS-giant binary system or large uncertainties in the observables for faint stars (see App.~\ref{app:perform_sim_clusters}). The box in the bottom right panel of Fig.~\ref{fig:clus_performance_sbi} shows a few simulated binary systems with relatively low mass-ratios predicted as single stars. This is also the case for the stars that were simulated to be single but were predicted as unresolved binaries with intermediate mass-ratios. Thus, this depicts the limited mass-ratio sensitivity of our method and that we can reliably predict the mass-ratios for only certain unresolved binaries that have $q_{\mathrm{true}}$ above a certain threshold, say $q_{\mathrm{thresh}}$. We compute this optimal $q_{\mathrm{thresh}}$ using the receiver operating characteristic curve (ROC), assuming that we can only detect binaries with $q>0.2$, and indicate this by the black dashed line in the figure. This mass-ratio threshold, $q_{\mathrm{thresh}}$, therefore depends on the cluster parameters and the observables used to obtain the posteriors\footnote{though it will also depend on $M_\mathrm{A}$, we opt to keep things simple.}. A detailed analysis on the completeness and contamination for a sample of simulated clusters is provided in App.~\ref{app:tpr_fpr_above_q_thresh} along with the efficacy of our iterative approach in App.~\ref{app:perform_iterative_cluster_param_method}. \referee{Further, we investigate the variation of $q$ residuals and uncertainties with magnitude and the simulated mass-ratio in App.~\ref{app:q_residual_abs_g_mag_true_q}.}

%-------------------------------------------------------------------

\section{Results for {\it Gaia} DR3 clusters}\label{sec:results}

We apply our method of obtaining of joint posterior distributions of six astrophysical model parameters for each subset of member stars described in Fig.~\ref{fig:g_hist_surveys} of 42 OCs with decent quality CMDs for which our method is able to predict good cluster parameters that are well-suited for the characterization of unresolved binaries. The membership lists of the 42 OCs comprise 28\,073 stars, of which posterior samples could not be obtained for 58 (due to $G_{\mathrm{BP}}$ flux overestimation in faint sources; \citealt{Riello2021}), while 814 were flagged as ‘photometric outliers’ (Sect.~\ref{subsec:apply_cluster_priors}). The typical statistical uncertainties for the remaining 27\,201 members are 0.08~in~$q$ and $0.01\,\mathrm{M}_\odot$ in $M_{\mathrm{A}}$. Since each cluster has a different range of inferred stellar masses and mass-ratios (see Tab.~\ref{tab:cluster_summary}), it is difficult to homogenise the analysis and perform a population study for several clusters. Therefore, we provide a case study of two clusters: NGC~2360 and NGC~2682 (M67) before comparing our results with the literature. \referee{A more detailed analysis of the $M_\mathrm{A}$-q distribution in cluster population is planned for a future study}. The complete versions of Tables~\ref{tab:cluster_summary} and \ref{tab:dataset}, containing the inferred parameters for the 42 OCs and the results for individual member stars, respectively, are available via the CDS.

\subsection{Examples: NGC 2360 and NGC 2682}

\begin{figure}
    \centering
    \includegraphics[width=0.5\textwidth]{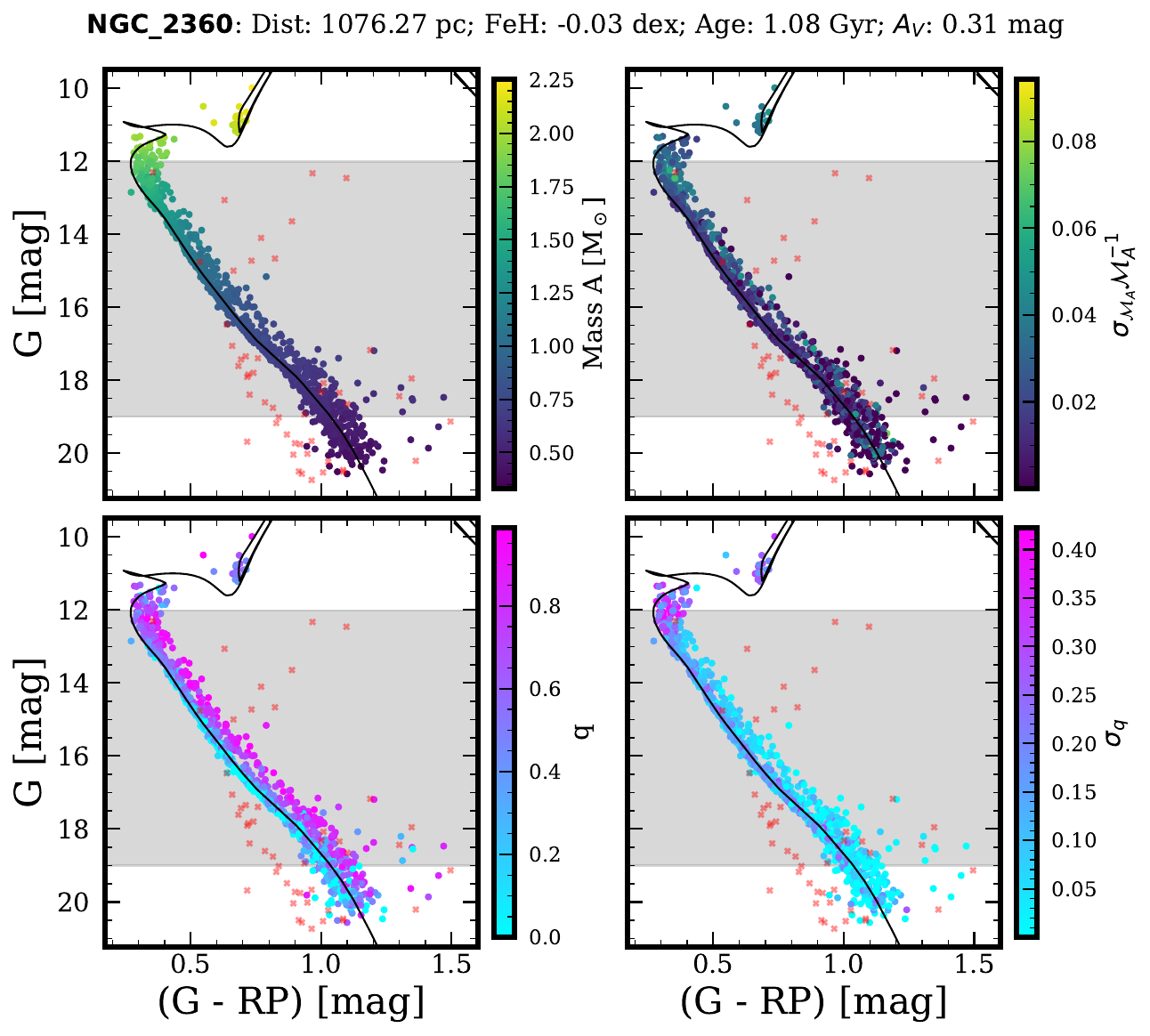}
    \includegraphics[width=0.5\textwidth]{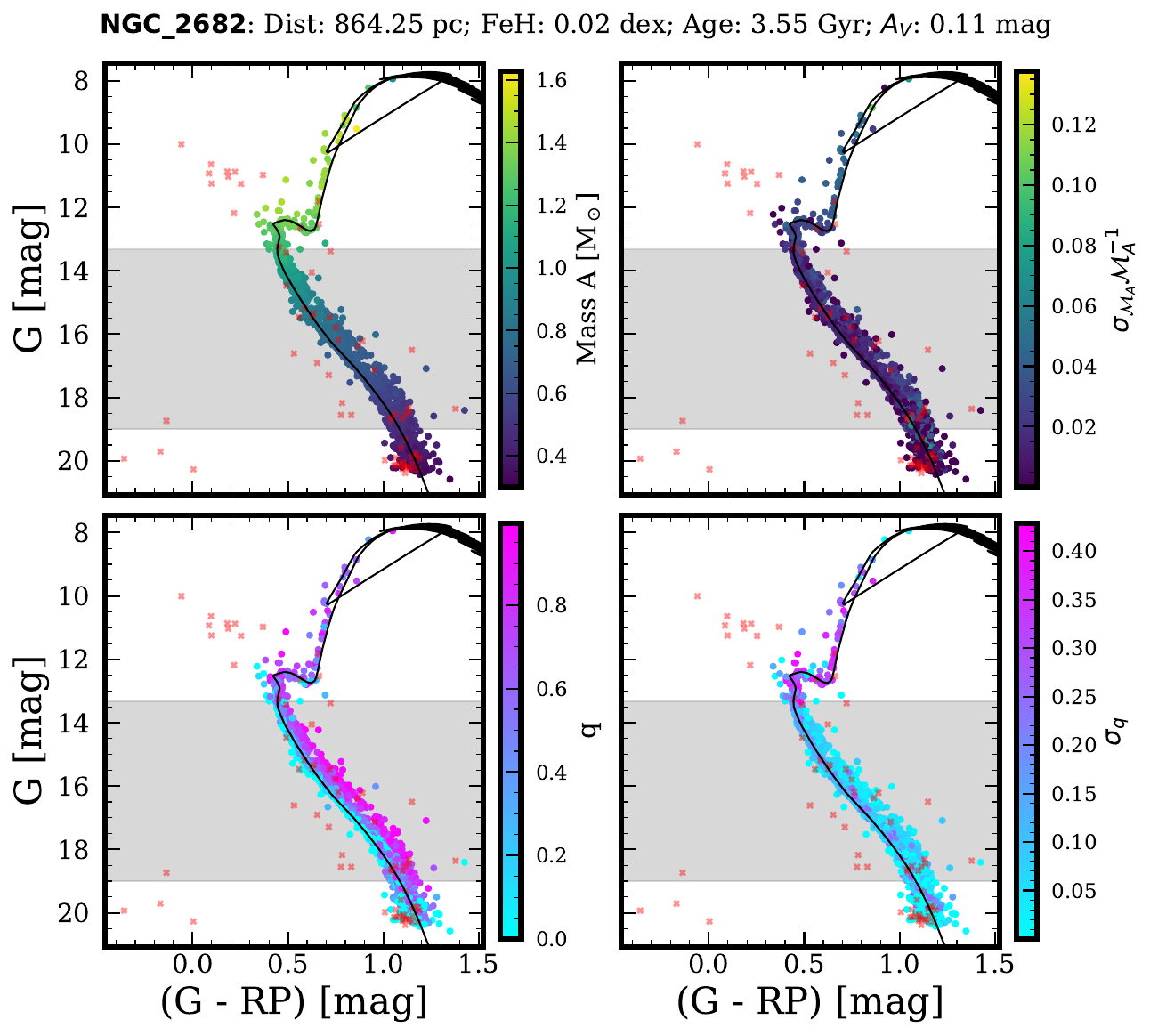}
    \caption{CMDs of NGC~2360 (\textit{top}) and NGC~2682~(M67) (\textit{bottom}). For each cluster, the CMDs are coloured by the primary masses and mass-ratios (left) and the corresponding uncertainties (right). The identified photometric outliers are shown by red crosses while the gray shaded region encloses the MS stars with $G \leq 19$ mag. The inferred cluster parameters, along with the spectroscopic metallicity used, are listed on top of each figure.}
    \label{fig:cmd_real_cluster}
\end{figure}

In Fig.~\ref{fig:cmd_real_cluster}, we show the CMDs of the two clusters with each star coloured by masses (top left) and mass-ratios (bottom left), along with their corresponding uncertainties (right panels). The solid black line is the PARSEC 1.2S isochrone corresponding to the estimated cluster parameters (provided on top of each subplot). For NGC~2360, most of the stars classified as photometric outliers (shown by red crosses) are likely field star interlopers, while for M67, these are mostly white dwarfs (WDs), blue stragglers, WD-MS binaries, or faint stars with poor $G_{\mathrm{BP}}$ photometry. Since modelling all these stellar populations in our simulation dataset is beyond the scope of this work, we only use the $q$ estimates of the MS stars in the gray-shaded region. Notably, the uncertainties in the mass-ratios increase significantly for stars close to the MSTO, which can be attributed to the increasing difficulty in distinguishing the single star and binary sequences with increasing primary mass where the two sequences intersect near the MSTO (\citealt{Hurley1998}; \referee{see also Fig.~\ref{fig:q_residual_vs_abs_g_true_q}}).

Through comparison with the simulated clusters, we obtain the mass-ratio threshold when using at least one of the 2MASS or WISE photometry along with \Gaia, $q_{\mathrm{thresh};\Gaia+}$ as $0.34$ and $0.25$ for NGC~2360 and M67~, respectively. On the other hand, the threshold when using only \Gaia photometry, $q_{\mathrm{thresh;\Gaia}}$, is similar for both clusters ($0.55$ and $0.56$, respectively). This demonstrates that using multi-band photometry (especially in the near/mid infrared) lowers the detection threshold of unresolved binaries, though the degree of improvements in detection capability varies across clusters.

\begin{figure}
    \centering
    \includegraphics[width=0.5\textwidth]{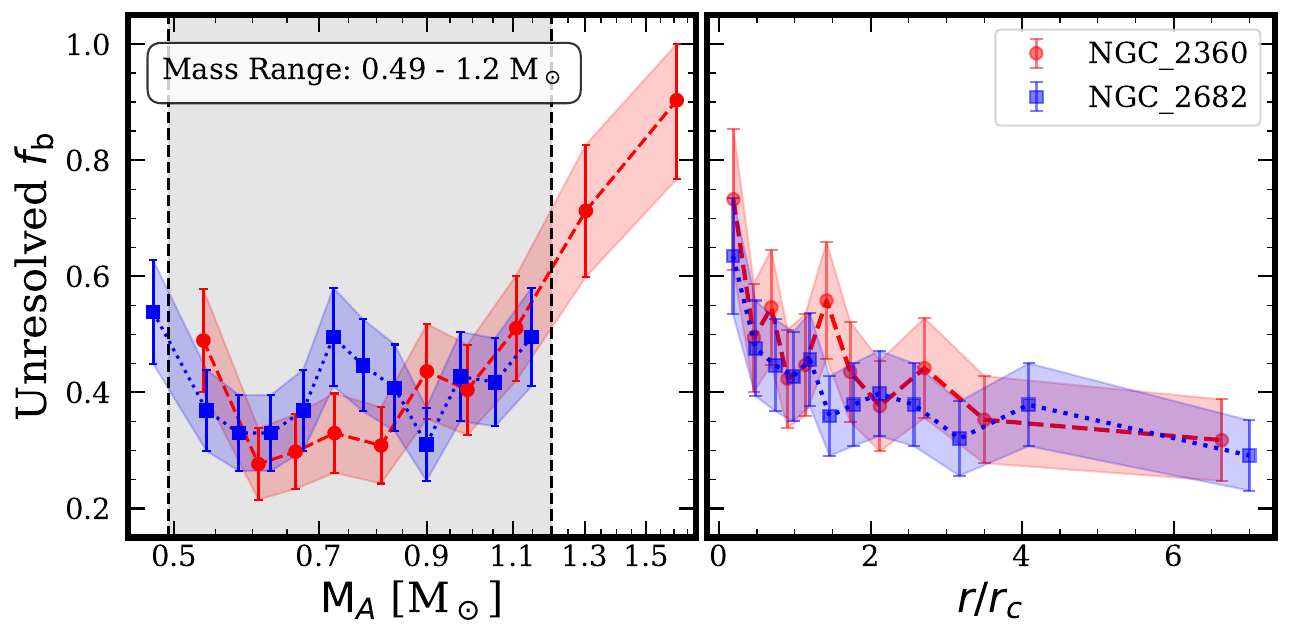}
    \caption{Unresolved binary fraction variation with the primary mass (left) and with the radial distance \referee{(adopted from \citetalias{Hunt2024})} from the cluster centre in units of $r_c$ (right). The gray shaded region enclosed by the dashed vertical lines indicates the mass range common between the two clusters.}
    \label{fig:bf_vs_mass_and_bf_vs_r_by_rc}
\end{figure}

In Fig.~\ref{fig:bf_vs_mass_and_bf_vs_r_by_rc} we plot the variation of the binary fraction with $M_\mathrm{A}$ and the radial distance \referee{(adopted from \citetalias{Hunt2024})} from the centre. The binary fraction in NGC~2360 increases monotonically for stars heavier than $\sim0.6\,\mathrm{M}_\odot$, similar to the trend observed in the Galactic field star population \citep{Offner2022}. In contrast, M67 shows noticeable dips in the binary fraction at about 0.6 and 0.9 $\mathrm{M}_\odot$. This variation can also be observed in previous studies such as \citet{Childs2025}. While these fluctuations may not be statistically significant (only at the $\sim 1.5\sigma$ level), it is likely that their origin is related to the previously observed gaps in the MS of OCs \citep{Balaguer-Nunez2005}. The right panel of Fig.~\ref{fig:bf_vs_mass_and_bf_vs_r_by_rc} shows the declining binary fractions with increasing radial distance from the cluster centre. This trend has been reported in several clusters \citep{Dalessandro2011, Giesers2019, Zwicker2024} and is an expected consequence of mass segregation (binaries tend to be more massive than single stars).

\begin{figure}
    \centering
    \includegraphics[width=0.42\textwidth]{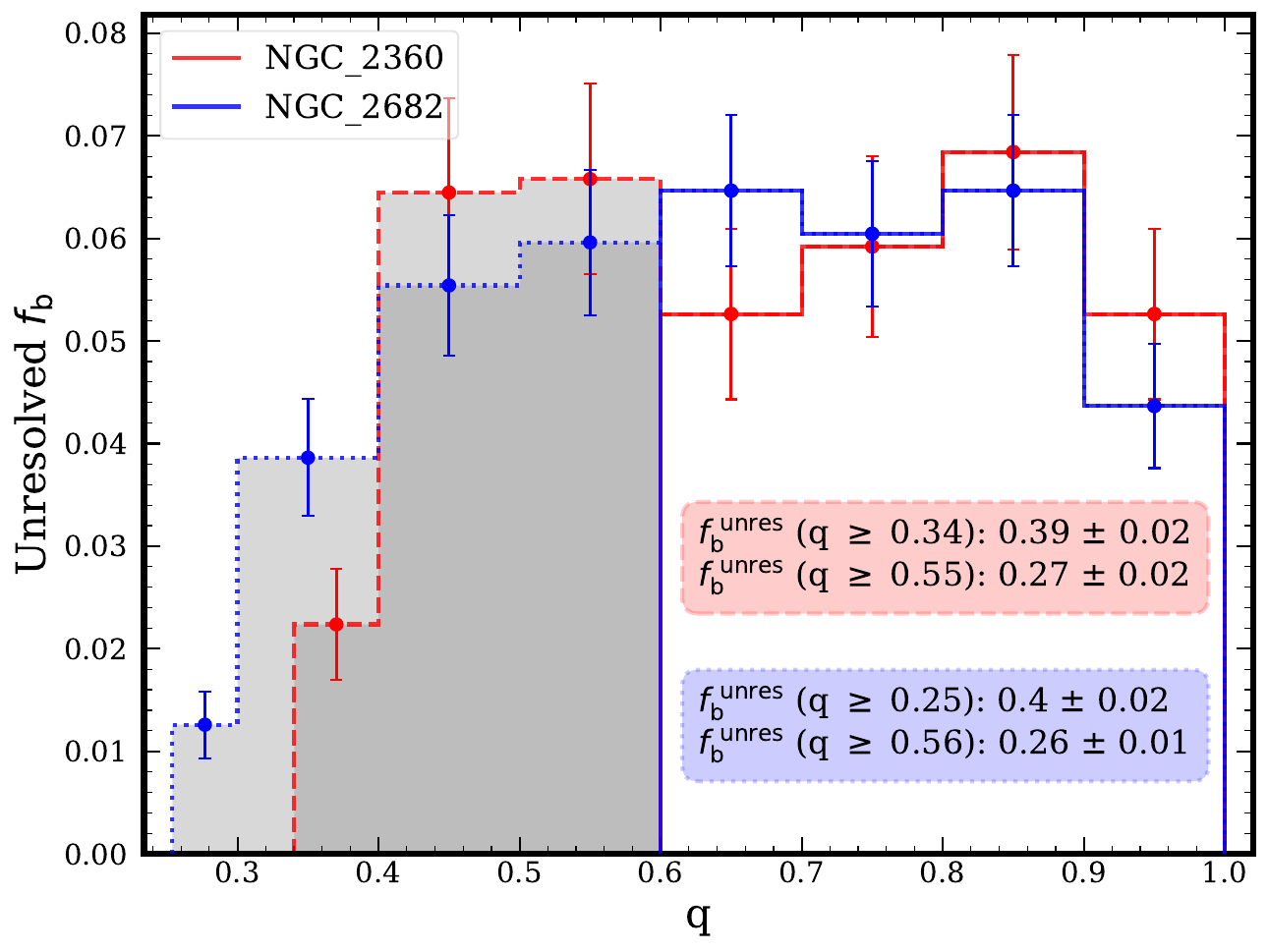}
    \caption{$q$ distributions of the detected unresolved binaries (in the mass range of 0.49 - 1.2 $\mathrm{M}_\odot$) in NGC~2360 and NGC~2682~(M67) shown by red and blue lines, respectively. The gray shaded region indicates the mass-ratio bins below $q_{\mathrm{thresh;\Gaia}}$ and the estimated binary fractions for the two mass-ratio thresholds viz., $q_{\mathrm{thresh;\Gaia}}$ and $q_{\mathrm{thresh;\Gaia}+}$  are reported along with Poisson uncertainties.}
    \label{fig:q_distribution_real_clus}
\end{figure}

To compare the mass-ratio distributions of the two clusters, we consider only the primary masses in the common mass range of 0.49 - 1.2 $\mathrm{M}_\odot$ (shaded area in Fig.~\ref{fig:bf_vs_mass_and_bf_vs_r_by_rc}) and plot their mass-ratio distributions in Fig.~\ref{fig:q_distribution_real_clus}. The gray-shaded regions represent the mass-ratio bins below $q_{\mathrm{thresh;\Gaia}}$ and are likely to be incomplete. We observe that the two distributions are very similar (all bins with $q>0.4$ are compatible within 1$\sigma$) and both exhibit a slight deficiency in the equal-mass binaries. \referee{This contrasts with observational studies of field binaries, which typically report an excess of equal-mass “twin” systems ($q \approx 1$; \citealt{Lucy1979, Tokovinin2000, El-Badry2019_twinexcess, Offner2022}). As shown in Fig.~\ref{fig:q_residual_vs_abs_g_true_q}, this discrepancy likely arises because the mode of the $q$ posterior distribution, truncated at $q=1.0$, is systematically biased toward lower values. Furthermore, the degeneracy between $M_\mathrm{A}$ and $q$ for high mass-ratio systems (see Fig.~1 of \citealt{Wallace2024}) can also introduce additional biases in our $q$ estimates. Since we do not impose any prior on the mass function that might partially alleviate this degeneracy, the apparent suppression of equal-mass binaries in our results is most likely a methodological artefact rather than a physical effect. We discuss some implications of this in Sect.~\ref{sec:discussion} and App.~\ref{app:q_residual_abs_g_mag_true_q}.} The binary fractions for $q \geq q_{\mathrm{thresh;\Gaia}+}$ and $q_{\mathrm{thresh;\Gaia}}$ obtained as $n_{binaries} (q \geq q_{\mathrm{thresh}})/n_{MS}$ for the two clusters are shown in the red and the blue boxes respectively where the uncertainties are estimated as Poisson errors i.e. $\sqrt{n_{binaries} (q \geq q_{\mathrm{thresh}}})/n_{MS}$, with $n_{MS}$ denoting the number of MS stars in a particular mass bin.

\subsection{Unresolved binary fractions of the 42 OCs}

\begin{figure}
    \centering
    % \sidecaption
    \includegraphics[width=0.5\textwidth]{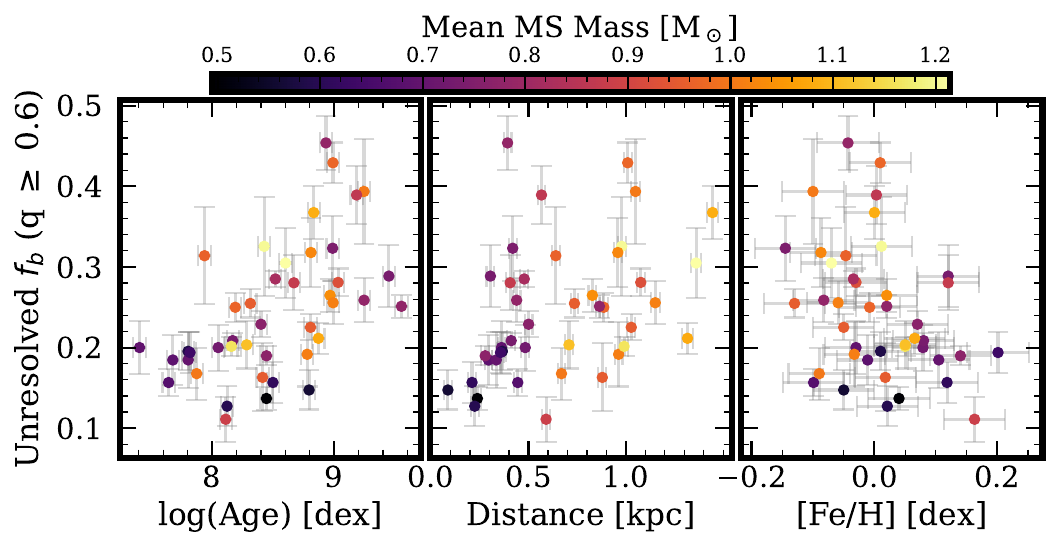}
    \caption{Dependence of the high mass-ratio ($q \geq 0.6$) unresolved binary fraction, $f_{b}$, with the cluster-wide parameters viz., log(Age) (\textit{Left}), distance (\textit{Middle}) and [Fe/H] (\textit{Right}) for the 42 OCs. The data points are coloured by the mean mass of MS primary stars. The gray error bars represent the Poisson uncertainties in the binary fractions and $1\sigma$ errors in the cluster parameters.}
    \label{fig:bf_trends}
\end{figure}

\begin{figure}
    \centering
    \includegraphics[width=0.46\textwidth]{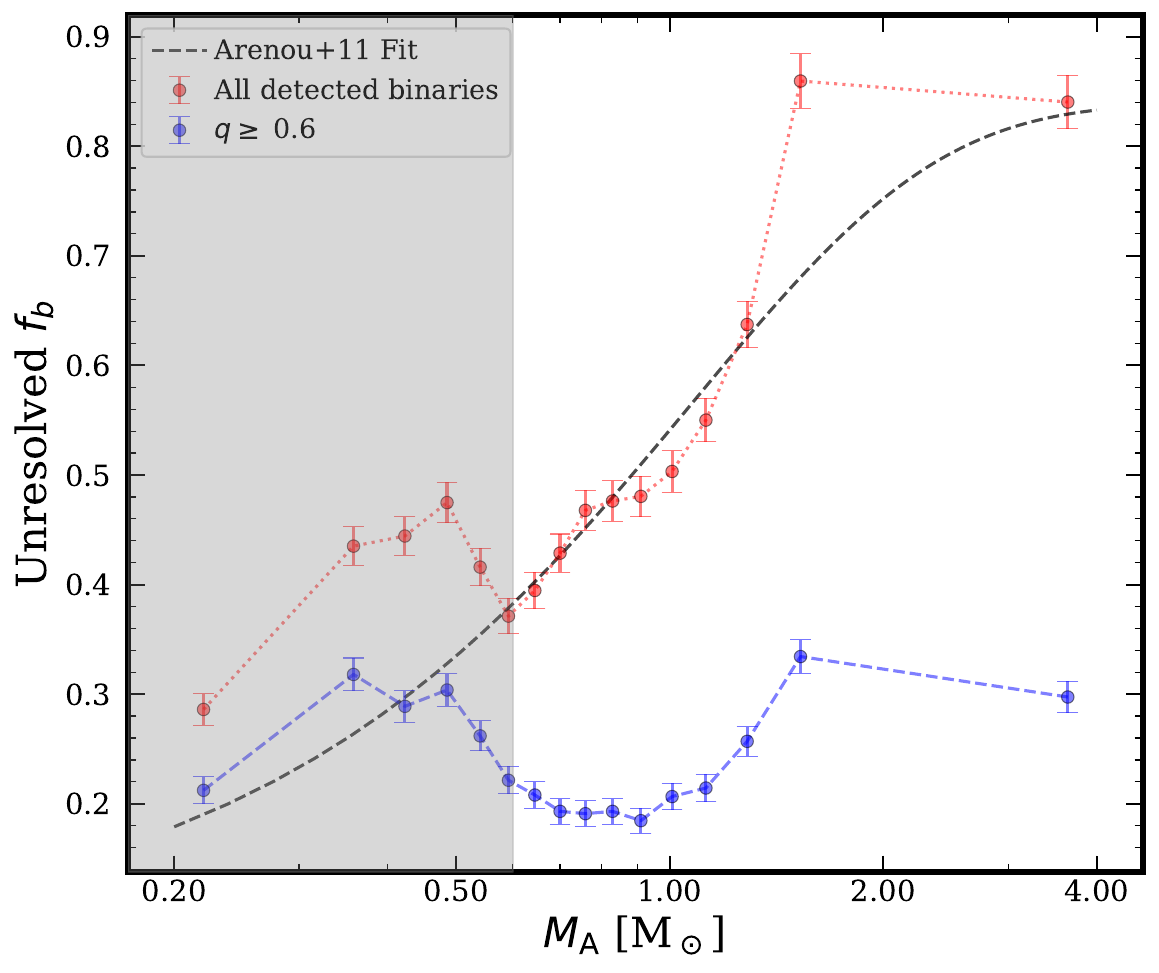}
    \caption{Unresolved binary fraction varying with the mass of primary star for all detected binaries (red) and only high mass-ratio binaries ($q \geq 0.6$; blue). The fit provided by \citet{Arenou2011} (Eq.~\ref{eq:binary_fraction_empirical}) to the observed variation in the field is overplotted by black dashed line. The shaded region corresponds to the masses $\leq 0.6\,\mathrm{M}_\odot$ wherein the binary fractions are expected to be spuriously inflated due to colour deviations of the observed \Gaia DR3 photometry from the PARSEC isochrones.}
    \label{fig:binary_vs_mass_stacked}
\end{figure}

Since our 42 analysed clusters have significantly different mass-ratio thresholds ranging from 0.2 to 0.56, we only consider binaries with mass-ratios greater than 0.6 in order to minimize the incompleteness effects in studying the dependence of binary fractions on fundamental cluster parameters such as age, metallicity and distance. Figure~\ref{fig:bf_trends} shows these trends, with the data points being coloured by the mean mass of the MS cluster members. The corresponding Kendall correlation matrix is shown in Fig.~\ref{fig:bf_corr_matrix}. The high mass-ratio binary fraction is correlated with the cluster ages, while it shows a weak correlation and anti-correlation with distance and metallicity, respectively. The correlation with the cluster age can be interpreted as a result of mass segregation and selective mass loss in the form of low-mass stars, which are less likely to host binary companions \citep{Moe2017, Winters2019}, thereby resulting in an increase in the fraction of stars in binary systems over time. The trend with the distance is also expected due to the higher number of close binaries that remain unresolved by \Gaia at larger distances (e.g. \citealt{Donada2023}). This is also consistent with the fact that we can only observe the brighter portion of the MS of distant clusters, thus mirroring the observed trend of increasing binary fraction with the mass of the primary star (Fig.~\ref{fig:binary_vs_mass_stacked}, Fig.~\ref{fig:bf_corr_matrix}). Furthermore, we observe a weak declining trend of the high mass-ratio binary fraction with the cluster metallicity, which agrees with the literature \citep[e.g.][]{Gao2017, Moe2019, Donada2023} but could also be explained by the apparent weak anti-correlation of metallicity with the distance for our sample of OCs (see Fig.~\ref{fig:bf_corr_matrix}). Therefore, a wider coverage in metallicity is required for clusters at similar ages and distances to corroborate this correlation.

Another trend that is widely studied, especially in the case of field star population, is the variation of binary fraction with the primary stellar mass \citep[e.g.][]{Li2022, Offner2022}. We explore this in Fig.~\ref{fig:binary_vs_mass_stacked} by plotting the unresolved binary fraction for both, all detected and only high mass-ratio binaries ($q \geq 0.6$) in red and blue, respectively, with each mass bin containing about 1400 MS stars. We also overplot the fit proposed by \citet{Arenou2011} (Eq.~\ref{eq:binary_fraction_empirical}) to the observed trend in the field star population. The fit for the Galactic field agrees well with our results for all detected binaries in OCs, specifically for stars heavier than 0.6~$\mathrm{M}_\odot$ (unshaded region in the plot). In addition, the high mass-ratio binary fraction shows a monotonic increase with the stellar mass for a similar mass range. However, the apparent decline in the binary fraction for 0.5–0.6~$\mathrm{M}_\odot$\footnote{We note that 90\% of the OCs in our sample have their lightest stars (with $G \leq 19$ mag) lighter than $0.5\,\mathrm{M}_\odot$} and an excess of binaries for lower stellar masses in both cases might be due to the well-known deviations of theoretical isochrones with observations at the low-mass regime \citep[e.g.][]{Bell2014, Fritzewski2019, Brandner2023, Wang2025}. In such cases, the modelled isochrone tends to be bluer (redder) than the observations, thereby resulting in a higher (lower) inferred binary fraction among low-mass stars (this is particularly the case for NGC~2287 and NGC~3532). As mentioned in Sects.~\ref{subsec:simulate_training_data} and \ref{sec:sim_clus_performance}, we did not model this effect in our simulations and therefore the binary fraction estimates in this mass regime may be biased. Future improvements to the synthetic photometry--either through empirical corrections or better low-mass stellar models--are necessary to mitigate this discrepancy and improve the reliability of binary identification in the low-mass regime.

%--------------------------------------------------------------------

\section{Comparison with the literature}\label{sec:comp_with_lit}

\subsection{Stellar masses}

\begin{figure}
    \centering
    \includegraphics[width=0.46\textwidth]{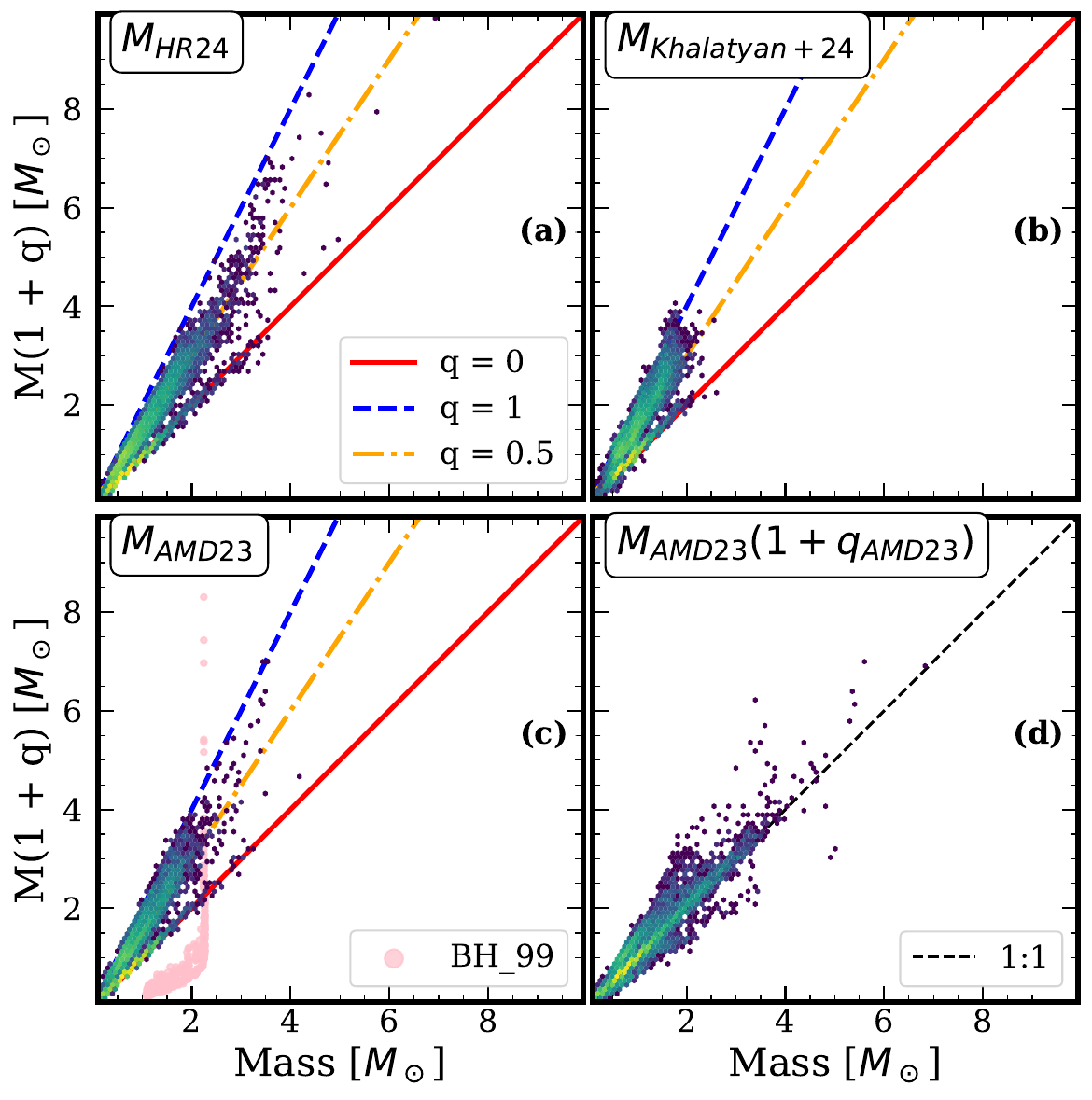}
    \caption{Comparison of the total mass of unresolved binaries with the stellar masses of the cluster members provided in the literature. \textit{(a)}: \citetalias{Hunt2024}; \textit{(b)}: \citet{Khalatyan2024}; \textit{(c)} and \textit{(d)}: Individual and total stellar masses i.e. including the masses of companions provided by \citetalias{Almeida2023}. The red (solid), orange (dash-dot) and the blue (dashed) lines indicate the $q=0$, $q=0.5$ and $q=1$ relations, respectively. The black dashed line in \textit{(d)} is the 1-1 line, whereas the pink circles in \textit{(c)} are the inconsistent \citetalias{Almeida2023} stellar masses reported for BH~99 members (see text for details).}
    \label{fig:stellar_mass_comp}
\end{figure}

Figure~\ref{fig:stellar_mass_comp} shows the comparison of the total mass of unresolved binary systems with the stellar masses provided by \citetalias{Hunt2024}, \citet[]{Khalatyan2024}, and \citet[][hereafter AMD23]{Almeida2023}. \citetalias{Hunt2024} and \citet{Khalatyan2024} use general-purpose approaches that do not take into account the stellar multiplicity. \citetalias{Hunt2024} used the $G$ magnitude, along with the $G_{\rm BP}-G_{\rm RP}$ colour indices for old clusters, to estimate stellar masses from PARSEC isochrones, having previously determined age, distance, extinction, and assuming solar metallicity. \citet{Khalatyan2024} trained the extreme gradient-boosted tree system, \texttt{xgboost} \citep{Chen2016}, on ground-based spectroscopic surveys \citep{Queiroz2023} to predict stellar masses of all $\sim 220$ million \Gaia DR3 stars with low-resolution BP/RP spectra \citep{DeAngeli2023}, most of them being field stars. In contrast, \citetalias{Almeida2023} specifically focus on OCs including unresolved binaries: they estimate the individual stellar masses in a Monte Carlo framework while comparing observed and simulated clusters, where the synthetic photometry is generated from PARSEC stellar models. We observe in Fig.~\ref{fig:stellar_mass_comp} that our masses for the single star systems agree well with the literature with the total mass of binary systems lying within the region enclosed by blue (dashed) and red (solid) lines, indicating $q=1$ and $q=0$, respectively. The exception is the case of BH~99 stellar masses reported by \citetalias{Almeida2023} in the panel (c) of the figure, shown by the pink data points. Since we do not see such discrepancy for the BH~99 cluster members in comparison with \citetalias{Hunt2024} and \citet{Khalatyan2024}, we conclude that it is the result of an inconsistency in \citetalias{Almeida2023} masses for this cluster, and consequently its mass-ratios, and remove it in the subsequent analysis. The comparison of the total stellar masses with \citetalias{Almeida2023} in panel (d) shows a general agreement between the two studies, with a median fractional absolute difference of $5^{+22}_{-4}~\%$.

\subsection{Binary mass-ratios and binary fractions}

\begin{figure}
    \centering
    \includegraphics[width=0.5\textwidth]{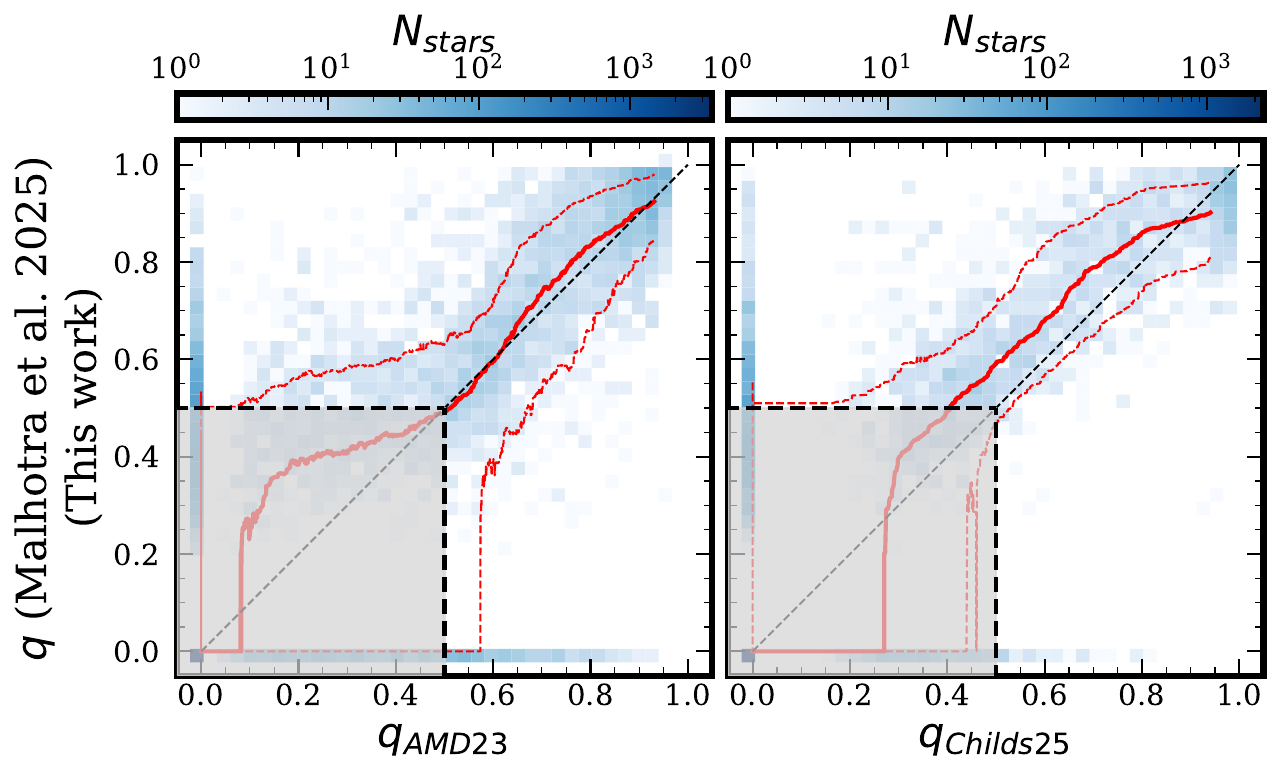}
    \caption{Comparison of the predicted mass-ratios with estimations by \citetalias{Almeida2023} (\textit{Left}) and  \citetalias{Childs2025} (\textit{Right}). The shaded region indicates low-$q$ region while the red solid and dashed lines denote the running median and the 16th-84th percentiles, respectively.}
    \label{fig:q_comp_lit}
\end{figure}

In Fig.~\ref{fig:q_comp_lit}, we compare our predicted mass-ratios with \citetalias{Almeida2023} and \citet[][hereafter Childs25]{Childs2025} for 30 (8342~members) and 7 (4890 members) overlapping OCs, respectively. The common low-mass-ratio ($q \leq0.5$) region is shaded gray, since the reliability of comparisons between different studies diminishes in this low-$q$ regime (e.g. \citetalias{Childs2025} found that their mass-ratios are only reliable for $q>0.5$). We find that there is a good agreement between our predictions and the previous studies, with a few ambiguously classified binaries and singles compared to \citetalias{Almeida2023} and our mass-ratios being typically $0.1$ higher compared to \citetalias{Childs2025}. While \citetalias{Almeida2023} rely only on the \Gaia DR2 photometry, we make use of the multi-band photometry including the near and mid infrared, leading to a better identification of lower-$q$ unresolved binaries. The differences in our comparison with \citetalias{Childs2025} are correlated with offsets in the metallicity (our study adopts spectroscopic [Fe/H] values from the literature, whereas \citetalias{Childs2025} inferred [Fe/H] from cluster CMDs) and derived extinctions. As shown in Fig.~\ref{fig:mae_vs_offset_in_clus_param}, an offset of about 0.1 dex in [Fe/H] and 0.1 mag in $A_V$ can result in significant differences in the predicted mass-ratios. A different photometric survey of Pan-STARRS1 used by \citetalias{Childs2025}, in contrast to the WISE survey used in our study, can also contribute to minor differences in the predicted mass-ratios. Nevertheless, we show that our predicted mass-ratios fare well when compared to the direct observations of SB2 binaries in M67 and NGC~2516 in Fig.~\ref{fig:q_comp_wocs}, taken from \citet{Geller2021} and \citet{Lipartito2021} respectively (crossmatched within 0.5"), where our mass-ratio predictions are slightly underestimated by a median absolute offset of 0.049 for M67 members.

\begin{figure}
    \centering
    \includegraphics[width=0.4\textwidth]{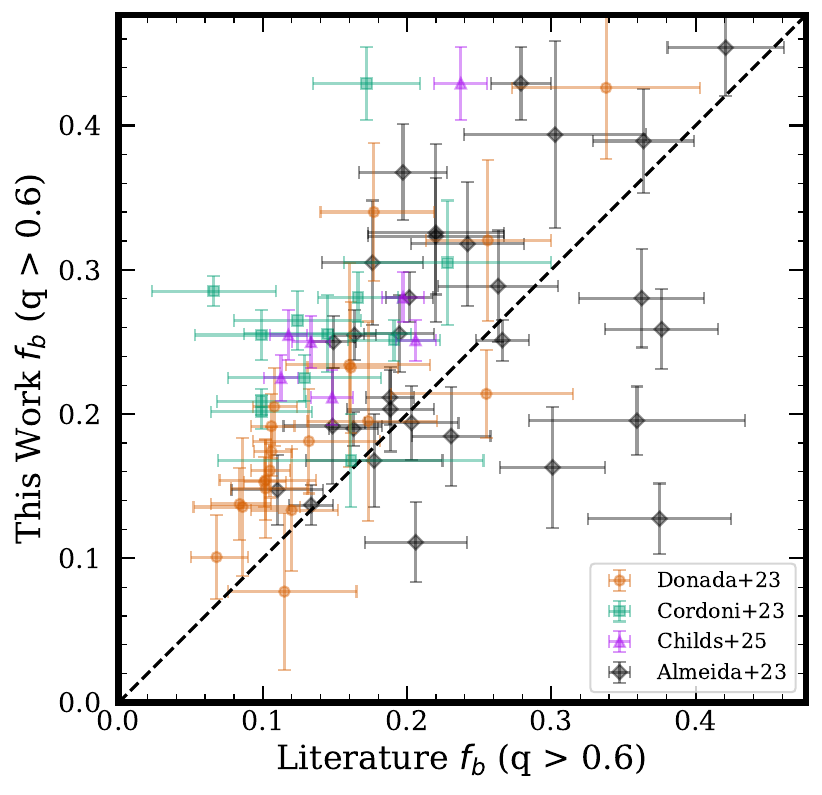}
    \caption{Comparison of high mass-ratio ($q > 0.6$) binary fractions with literature while using the same mass range, when available.}
    \label{fig:binary_frac_comp}
\end{figure}

The comparison of recovered high $q$ unresolved binary fractions with a few previous studies (\citealt{Donada2023, Cordoni2023}; \citetalias{Almeida2023, Childs2025}) is shown in Fig.~\ref{fig:binary_frac_comp}, considering the overlapping stellar mass range, wherever such data are available. We find that, in general, our binary fractions are slightly higher (by a factor of $\sim1.4$) than the other studies that, except for \citetalias{Childs2025}, use only the \Gaia photometry. Using a wider wavelength range while including infrared photometry along with the measurements in the optical band has been shown to result in a better detection rate of unresolved binaries in comparison to optical photometry \citep[e.g.][]{Steele2011, Malofeeva2022}. However, binary fractions larger than those reported by \citetalias{Childs2025} may arise from differences in the membership lists as well as the slightly higher mass-ratios predicted by our method compared to BASE-9 \citep{Robinson2016} used by \citetalias{Childs2025}.

\subsection{Fundamental cluster parameters}\label{subsec:comp_clus_param_with_lit}

\begin{figure*}
    \centering
    % \sidecaption
    \includegraphics[width=0.9\textwidth]{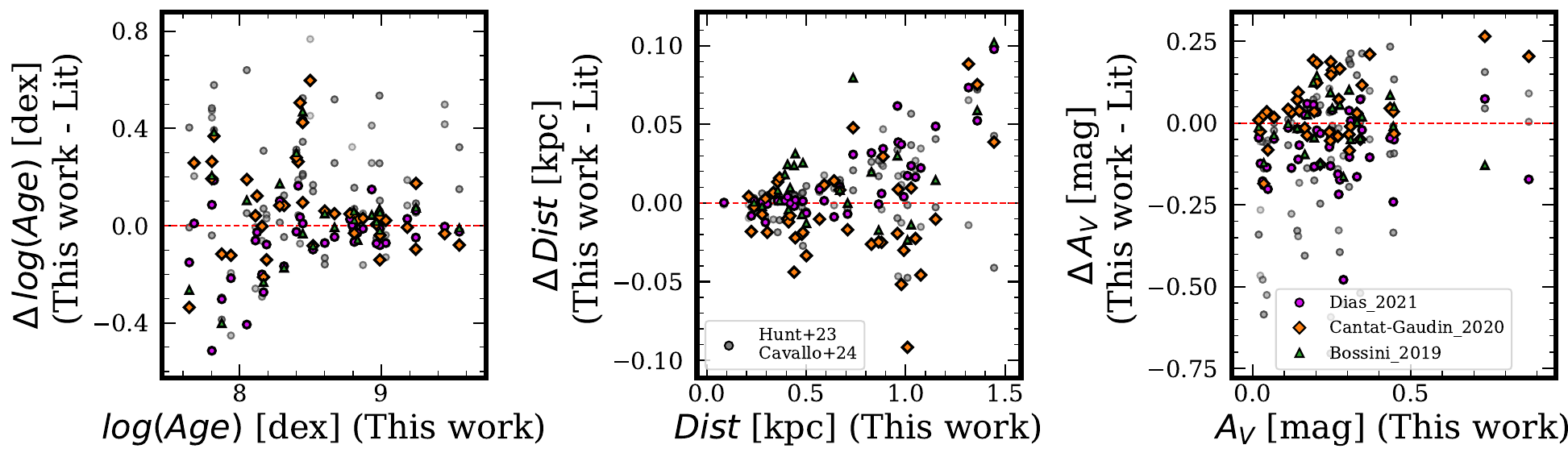}
    \caption{Comparison of cluster parameters with the literature where the coloured data points denote the estimates from \citealt{Bossini2019, Cantat-Gaudin2020, Dias2021} and the gray circles indicate the parameters inferred by \citealt{Hunt2023} and \citealt{Cavallo2024}. The red dashed line indicates the line of ideal agreement.}
    \label{fig:cluster_param_comp_with_lit}
\end{figure*}

Finally, we compare our inferred cluster parameters (log(Age), distance, and $A_V$) with some well-known cluster census studies \citep{Bossini2019, Cantat-Gaudin2020, Dias2021, Hunt2023, Cavallo2024} in Fig.~\ref{fig:cluster_param_comp_with_lit}. Overall, our results show a broad agreement with previous studies, with a few deviations in the ages of young OCs and in the distance estimates of farther OCs. Only the $A_{v}$ values display larger offsets relative to the estimates of \citet{Hunt2023} and \citet{Cavallo2024}. We attribute this to the inclusion of infrared photometry in our analysis, which helps to partially break the age–extinction degeneracy—a common limitation in isochrone fitting methods. The typical offsets in comparison to \citet{Bossini2019, Cantat-Gaudin2020, Dias2021} are: $0.08^{+0.19}_{-0.05}~\mathrm{dex}$ for log(Age), $0.01^{+0.02}_{-0.01}~\mathrm{kpc}$ for distance, and $0.06^{+0.11}_{-0.03}~\mathrm{mag}$ for $A_V$. Some well-known young OCs, such as Blanco~1, have ages slightly overestimated by 0.2-0.4 dex, but we show in Fig.~\ref{fig:mae_vs_offset_in_clus_param} that such an offset in the OC age does not affect our inferred primary stellar masses and mass-ratios significantly.

%--------------------------------------------------------------------

\section{Discussion}\label{sec:discussion}

In this work, we have introduced a new, fast strategy for characterizing unresolved binaries in OCs using Simulation-Based Inference (SBI), while simultaneously fitting for the fundamental cluster parameters. We successfully applied our method on a sample of 42 OCs yielding typical uncertainties of 0.08 in $q$ and 0.01 $\mathrm{M}_\odot$ (or 1\%) in $M_{\mathrm{A}}$. The principal objectives of this study were to: (i)~explore the trends of binary fraction variation in different clusters, (ii)~compare the binary stellar population in the OCs with that in the Galactic field, and (iii)~provide a critical assessment of the performance and limitations of our method.
We expand on our findings and place them in a broader context in the following sections.

\subsection{Binary fraction trends: Dependence on age and [Fe/H]}\label{subsec:discuss_bf_trends}

The dependence of the binary fraction, both in the Galactic field and in open clusters, on age and metallicity has been explored extensively in previous studies \citep[e.g.,][]{El-Badry2019, Niu2020, Hwang2021, Donada2023, Pang2023, Alexander2025}. In Fig.~\ref{fig:bf_trends}, we find that the high mass-ratio unresolved binary fraction in OCs tends to increase with age. This trend has previously been reported \citep{Jadhav2021, Donada2023, Childs2025}, and specifically for cluster cores by \citet{Cordoni2023}, whose findings also reflect the effects of mass segregation. While \citet{Niu2020} do not identify a clear correlation, they suggest that clusters with larger evolutionary parameters may exhibit higher binary fractions. Conversely, \citet{Pang2023} and \citet{Thompson2021} report a declining trend in the overall binary fraction with cluster age. Specifically, \citet{Pang2023}, selecting binaries with $q > 0.4$ using a ridge-line method, note a modest increase for older clusters beyond 100~Myr, whereas \citet{Thompson2021}, based on a sample of only eight OCs, find that the binary fraction decreases during the first 200~Myr and becomes constant for older OCs. While using probabilistic generative modelling approach, \citet{Alexander2025} reported a decline in the binary fraction with cluster age. However, N-body simulations by \citet{PortegiesZwart2001} and \citet{Fregeau2009} suggest that the initial decrease arises from the destruction of soft binaries, while a subsequent increase in binary fraction occurs due to the preferential evaporation of low mass single stars. Considering that most of the clusters in our sample are older than 100~Myr, the observed increase in high mass-ratio binary fraction is likely a consequence of the latter process while confirming the early-phase decline in binary fraction predicted by simulations would require a larger sample of younger OCs.

With regards to the metallicity dependence of binary fraction, previous studies \citep{Grether2007, Yuan2015, Badenes2018, Moe2019, Mazzola2020} have reported an anti-correlation for close binaries in the field star population. \citet{El-Badry2019} further found that such an anti-correlation emerges only for binaries with separations $\lesssim$ 200 AU. While studying wide binaries using metallicities from LAMOST and \Gaia astrometry, \citet{Hwang2021} observed that the fraction of wide binaries first increases with metallicity, depicts a maximum close to [Fe/H] $\sim$ 0 dex along with a sharp decline for higher metallicities. Since the wide binary population is rare in large OCs \citep{Scally1999, Parker2009, Deacon2020, Rozner2023, Cournoyer-Cloutier2024}, we can expect differences in their results and the right panel of Fig.~\ref{fig:bf_trends}. Given that OCs probe a smaller range of metallicities compared to field stars \citep{Spina2022}, previous studies \citep[e.g.][]{Donada2023, Cordoni2023, Childs2024} have reported only a weak or no correlation with the metallicity. Similar conclusions have also been drawn for globular clusters \citep{Milone2012, Ji2015}. Although a declining trend of the binary fraction with OC metallicity is observed in Fig.~\ref{fig:bf_trends}, sample selection effects (see Fig.~\ref{fig:bf_corr_matrix}) make it difficult to robustly establish such a dependence. A larger sample including metal-poor OCs would be useful to better constrain this correlation.

\subsection{Binary fractions in OCs distinct from the Galactic field?}\label{subsec:discuss_bf_vs_mass_stacked}

Binary fractions of main-sequence stars, particularly FGK-type stars, have been found to be broadly consistent with those observed in OCs \citep{Bouvier1997, Patience2002, Deacon2020, Torres2021}. Recent studies (\citealt{Jadhav2021, Cordoni2023}; \citetalias{Childs2025}) have investigated the dependence of multiplicity on the primary mass using \Gaia DR2/DR3 OC data. \citet{Cordoni2023} reported that, among 78 OCs, 72 exhibit a relatively flat binary fraction distribution with stellar mass. In contrast, \citet{Jadhav2021} and \citetalias{Childs2025} report a strong positive correlation between binary fraction and primary mass for stars $\gtrsim 0.7\,\mathrm{M}_\odot$, while lower-mass stars exhibit a declining trend, which \citetalias{Childs2025} attribute to an increased field-star contamination. In Fig.~\ref{fig:binary_vs_mass_stacked}, we combined all members of our 42 OCs to examine the global distribution of binary stars with the primary stellar mass. For stars heavier than $0.6\,\mathrm{M}_\odot$, the observed distribution agrees well with trends in the field population, whereas a higher multiplicity rate is observed among the lower masses. We attribute this primarily to the known discrepancies in PARSEC isochrones in the low-mass regime \citep{Wang2025}, where bluer isochrones can artificially increase the inferred unresolved binary fraction, while significant field star contamination is unlikely given our adopted $G$-band magnitude cut of $\lesssim19$ mag. Nevertheless, other explanations for the observed excess in low-mass binaries are also possible. One is that low-mass stars preferentially occur in high mass-ratio systems \citep[e.g.][]{Goodwin2013}, though opposing evidence has been reported \citep{Duchene2013}. Another possibility, supported by N-body simulations \citep[e.g.][]{PortegiesZwart2001, delaFuenteMarcos2010}, is the preferential loss of single low-mass stars, which can produce an apparent increase in the cluster binary fraction both with age (Fig.~\ref{fig:bf_trends}, Sect.~\ref{subsec:discuss_bf_trends}) and within the low-mass bin dominated by M-type stars. Resolving these effects in detail will require improved theoretical isochrone models in the low-mass regime and larger samples of OCs.

In summary, taking into account the evidence that most stars in the Galaxy form in compact aggregates \citep{Lada2003, Quintana2025}, even though most of them dissolve very quickly \citep{Lamers2005, Anders2021}, our analysis suggests that there are similarities among the binary stellar populations in OCs and the Galactic field. However, the literature shows that there are differences in the separation distributions that cannot be explained without the production of wide binaries in isolated star formation \citep{Parker2009} or a huge contribution from small clusters, in which wide binaries have been shown to have higher survival rates through N-body simulations \citep{Rozner2023}.

\subsection{Method performance and scope}\label{subsec:discuss_method_evaluation}

Our predicted primary stellar masses are in excellent agreement with previous studies and depict that, as expected, a significant fraction of the 200 million stars analysed by \citet{Khalatyan2024} are unresolved binaries. Depending on the light ratio, treating a binary system as a single star can strongly bias the derived stellar parameters, in particular stellar masses (e.g. \citealt{El-Badry2018}; \citealt{Anders2022} App. D; \citetalias{Almeida2023}). This highlights the importance of accounting for the presence of binary companions in future analyses to obtain more accurate stellar parameters. Figure~\ref{fig:mass_seg_dual_plot} serves as one such illustration in which the assumption of only single stars can lead to a markedly different interpretation of the dynamical state of a cluster.

With respect to computational cost, our method, once trained, is very fast compared to previous MCMC methods employed by \citetalias{Almeida2023} and \citetalias{Childs2025}. Using SBI, we are able to obtain 20\,000 posterior samples for model parameters within 2-3 seconds on a single core for one star while each iteration in the cluster parameter estimation loop takes about 8 seconds for a cluster with about 1800 member stars running on 7 cores. Moreover, as shown in depth in Sect.~\ref{sec:sim_clus_performance} and App.~\ref{app:perform_sim_clusters}, our method produces competitive estimates for masses and mass-ratios while inferring well-fitted isochrones to the observed CMD. 

Nevertheless, there are still some limitations to our method: (i)~the cluster parameter fitting iterative loop can struggle to estimate the parameters of young OCs and those affected by differential extinction, (ii)~the degeneracy in $M_\mathrm{A}$ and $q$ \referee{along with the mode of $q$ posterior biased towards lower values in} the high-$q$ regime can result in some underestimated mass-ratios of nearly-equal-mass binaries, (iii)~systematics in the stellar models (specifically for low-mass stars) likely result in unreliable characterization of low-mass binaries, and (iv)~triple and higher-order systems are not taken into account. We plan to implement future improvements in our iterative cluster parameter fitting loop to also include some of the other clusters that were classified as `Good' or `Poor' in the study of characterization of unresolved binaries. The mass-$q$ degeneracy can probably be addressed by adding a prior on the mass function in OCs. However, we refrain from applying such a prior in this first study because of the limited understanding of how the mass-function prior might influence the estimated total OC mass, which is beyond the scope of this work.
%-------------------------------------------------------------------

\section{Conclusions}\label{sec:conclusions}

In this paper we have outlined a new and efficient method for detecting and characterizing unresolved binaries in open clusters while simultaneously deriving optimal cluster parameters. The method combines multi-band photometry from \Gaia, 2MASS, and WISE with \Gaia parallaxes to infer posterior distributions of primary mass, mass-ratio, metallicity, age, distance, and line-of-sight extinction for individual stars using Simulation-Based Inference (SBI). Using an iterative framework to select member stars by their evolutionary phases, we robustly estimate the global cluster parameters while keeping the metallicity fixed to a spectroscopic estimate obtained from the literature.

From custom cluster simulations, we find that we can reliably detect unresolved binaries in OCs above mass-ratio thresholds ranging from $q_\mathrm{thresh}\sim0.2$–0.56, depending on cluster properties and the set of available observables. Analysing two example OCs, NGC~2360 and NGC~2682 (M67), we find that NGC~2360 shows a monotonic increase in the binary fraction for masses $\gtrsim0.6\,\mathrm{M}_\odot$, whereas M67 exhibits a number of dips, most likely due to gaps in the MS. We confirm that both clusters show higher binary fractions in the cluster cores. 

With this paper, we publish the results for 42 OCs for which our method works most successfully (OCs with clean and well-fitted CMDs suitable for studying their binary stellar populations) and provide a catalogue of individual stellar masses and mass-ratios for 27\,201 member stars with typical uncertainties of 0.08~in~$q$ and $0.01\,\mathrm{M}_\odot$ in $M_\mathrm{A}$. An overall analysis of the variation of the unresolved binary fraction with cluster-wide parameters reveals a positive correlation with cluster age and mean primary mass of MS stars and a weak anti-correlation with the cluster metallicity. The dependence of binary fractions on the primary stellar mass, when collating all cluster member stars, demonstrates a similar variation as observed in the Galactic field for stars heavier than $\gtrsim0.6\,\mathrm{M}_\odot$, while lower masses are likely affected by discrepancies in the stellar models we use.

\section*{Data availability}

Online versions of Tables~\ref{tab:cluster_summary} and \ref{tab:dataset} containing the cluster parameters for the 42 clusters and the results for individual cluster members, respectively, are are only available in electronic form at the CDS via anonymous ftp to \href{cdsarc.u-strasbg.fr}{cdsarc.u-strasbg.fr} (130.79.128.5) or via \href{http://cdsweb.u-strasbg.fr/cgi-bin/qcat?J/A+A/}{http://cdsweb.u-strasbg.fr/cgi-bin/qcat?J/A+A/}.

\begin{acknowledgements}
\referee{We thank the anonymous referee for constructive feedback that improved the manuscript.} The authors acknowledge all members of the GaiaUB group for insightful discussions throughout the course of this project. \referee{This work was (partially) supported by the Spanish MICIN/AEI/10.13039/501100011033 and by "ERDF A way of making Europe" by the European Union through grant PID2021-122842OB-C21 and PID2024-157964OB-C21, and the Institute of Cosmos Sciences University of Barcelona (ICCUB, Unidad de Excelencia Mar\'{\i}a de Maeztu) through grant CEX2024-001451-M and the project 2021-SGR-00679 GRC de l'Agència de Gestió d'Ajuts Universitaris i de Recerca (Generalitat de Catalunya).}. \referee{This research was partially funded by the Horizon Europe HORIZON-CL4-2023-SPACE-01-71 SPACIOUS project funded under Grant Agreement no. 101135205.} FA acknowledges financial support from MCIN/AEI/10.13039/501100011033 through grant RYC2021-031638-I, co-funded by European Union NextGenerationEU/PRTR.

This work has made use of data from the European Space Agency (ESA) mission {\it Gaia} (\url{https://www.cosmos.esa.int/gaia}), processed by the Gaia Data Processing and Analysis Consortium (DPAC, \url{https://www.cosmos.esa.int/web/gaia/dpac/consortium}). Funding for the DPAC has been provided by national institutions, in particular the institutions participating in the {\it Gaia} Multilateral Agreement. This research makes use of public auxiliary data provided by ESA/Gaia/DPAC/CU5 and prepared by Carine Babusiaux.
\end{acknowledgements}

\bibliographystyle{aa} % style aa.bst
\bibliography{biblio}

%-----------------------------------------------------------------

%------------------------------------------------------------------

\begin{appendix}

\section{Cross-match with 2MASS and WISE catalogues}\label{app:2mass_wise_crossmatch}

We use the DPAC crossmatch catalogues for 2MASS and allWISE to retrieve the corresponding photometry information for the \citetalias{Hunt2024} OC member stars from the respective survey catalogues. This is accomplished by crossmatching the \Gaia \texttt{source\_id} column with the crossmatch catalogues and using the corresponding \texttt{clean\_tmass\_psc\_xsc\_oid} and \texttt{allwise\_oid} columns in the DPAC curated catalogues\footnote{\href{https://gea.esac.esa.int/archive/documentation/GDR1/datamodel/Ch3/}{https://gea.esac.esa.int/archive/documentation/GDR1/datamodel/Ch3/}} for 2MASS and WISE surveys respectively. Due to different spatial resolutions of the three surveys, it is necessary to filter out the crossmatched sources that might be affected by blending issues due to presence of close neighbours, which might have been unresolved due to poorer effective spatial resolution of one survey. We apply the following quality cuts to only select the crossmatched stars in 2MASS and WISE having "good" photometry: For 2MASS, the photometric quality flag (ph\_qual) is set to "AAA" and "AA**" for WISE since only $W_1$ and $W_2$ are the photometric bands of interest. Additionally, for both 2MASS and WISE, the \texttt{number\_of\_mates} parameter is set to 0.

\begin{figure}[h]
    \centering
    \includegraphics[width=0.5\textwidth]{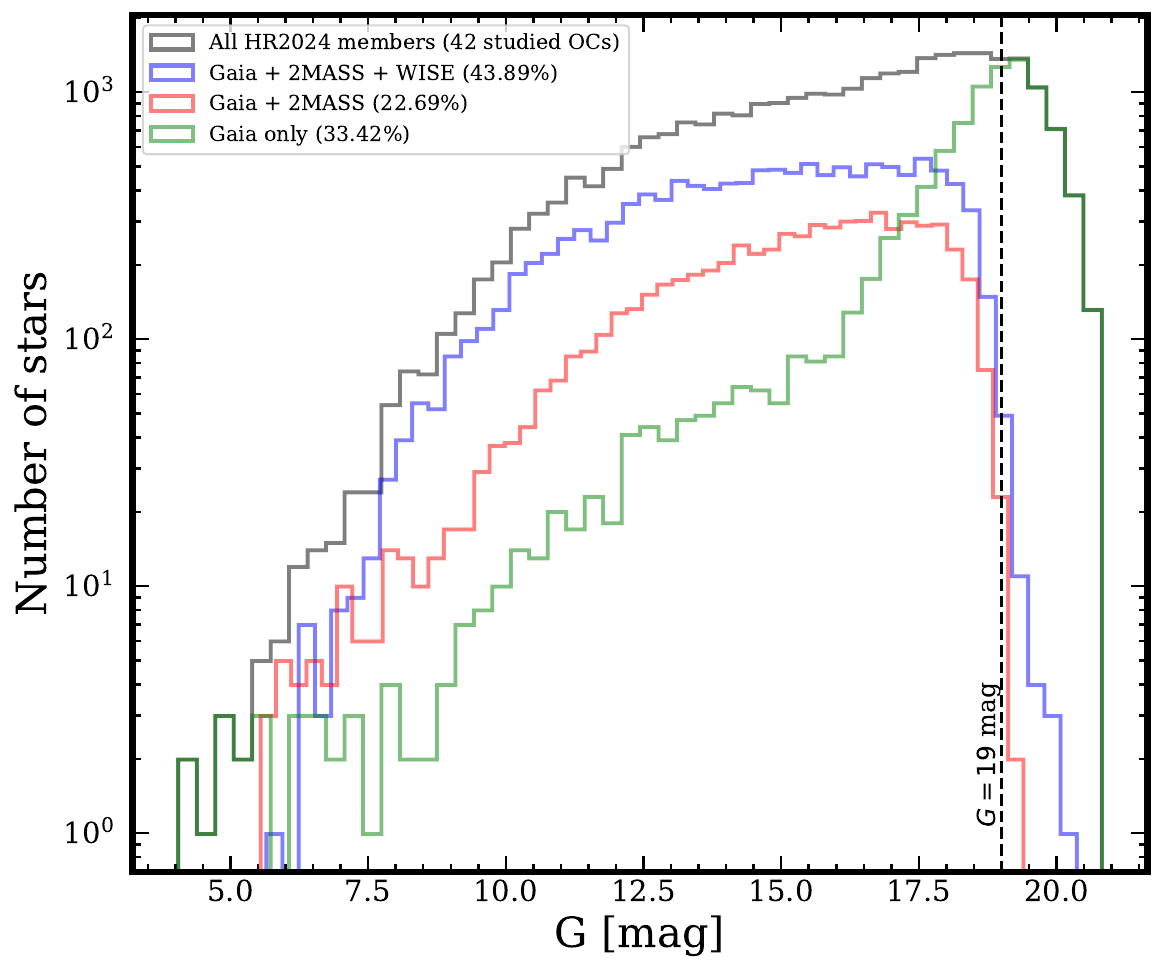}
    \caption{$G$ mag distribution of \citetalias{Hunt2024} member stars of 42 OCs divided into different sets based on the availability of good photometry (see text for more details) from 2MASS and WISE. The dashed vertical line marks $G = 19$ mag, beyond which very few stars have "good" photometric measurements.}
    \label{fig:g_hist_surveys}
\end{figure}

Figure~\ref{fig:g_hist_surveys} shows the distribution of the apparent $G$ magnitude of the members of the analysed 42 clusters (see Sect.~\ref{sec:results}) with different coloured histograms representing three subsets of stars based on the availability of good photometry. Therefore, most stars brighter than $G$ $\lesssim$ 17 mag have good photometry available from either of the infrared surveys, with \Gaia being the dominant source of photometry for fainter stars. We derive one SBI posterior for each subset of member stars.

\section{Estimation of cluster parameters}\label{app:cluster_param_iter_method}

We adopt an iterative procedure to estimate the cluster parameters as follows:
\begin{enumerate}
    \item We adopt the metallicity estimate from the literature, while for the rest of the parameters we take the mean value of the posterior samples of each star obtained using SBI (Fig.~\ref{fig:iterative_schematic}) with the resulting mean being considered as the initial guess for that parameter. These cluster parameter priors are then used to assign weights to each posterior sample.
    \item We only select the stars with at least 2\% valid samples (samples within 5$\sigma$ from the mean cluster parameter prior) while keeping the metallicity fixed and do the following:
    \begin{enumerate}
        \item log(Age): Since the age of a cluster is best estimated by the stars near the MSTO, especially for old OCs, we used at least 5 stars near the MSTO and computed the weighted mean and standard deviation (weighted by the individual errors; \citealt{Soubiran2018}) of their weighted log(Age) modes as the corresponding value for the next iteration. At each iteration, these stars are selected as being the nearest neighbours to the MSTO stars in the theoretical PARSEC isochrones. \referee{This is implemented by using the "label" provided with the PARSEC 1.2S isochrones that (approximately) denotes the evolutionary stage of each star. We assign these labels from the theoretical isochrones to the stars in the observed CMD using a nearest-neighbour algorithm and remove any star labelled as a MS or a pre-MS star while estimating the cluster age. Though this selection might not be completely accurate for the first few iterations, it improves significantly during the first 10 iterations (Fig.~\ref{fig:cluster_param_iter_method}). To also account for the case of young clusters with no or very few evolved stars, we do a check at the 10th iteration if that is the case and proceed to use stars in the upper and lower MS to automate the selection of suitable stars to estimate its age. We only use a vary small fraction of stars in the intermediate portion of the MS ($M_\mathrm{G} \sim 3.0 - 8.0$ mag) due to a largely indiscernible nature of isochrones in that region where the cluster ages in range $\sim 100 - 500$ Myr \citep{Rottensteiner2024} are highly degenerate. We test the robustness of this selection method in App.~\ref{app:perform_iterative_cluster_param_method}.}
        \item Distance and extinction: The distance and extinction priors for the next iteration are computed as the weighted mean and standard deviation of the weighted modes of the individual parameter values of each star brighter than $G = 18$ mag. After 10 iterations we do not include stars significantly displaced from the single star sequence for extinction estimation due to systematically overestimated extinctions for unresolved binaries.
    \end{enumerate}
    \item At each step, the error in the computed cluster parameter is bounded by a lower value of 0.05 dex for log(Age) and [Fe/H], 0.05 mag for $A_V$ and 2\% for distance. Although decreasing the lower bound of errors does not lead to significant improvements in the accuracy of the cluster parameters, it results in a large number of stars being classified as outliers due to the absence of a posterior sample within the 5$\sigma$ of the mean cluster parameter. However, this uncertainty likely represents only a lower bound, as systematic errors arising from the PARSEC stellar models and related factors are expected to dominate the uncertainties in our inferred cluster parameters.
    \item The method is considered to be converged if all the 6 metrics viz., the mean log(Age), distance, and extinction of the OC and their corresponding errors do not change by more than 0.1\% in the last 5 iterations. For some OCs, there can be instances where this criteria is not met even though the corresponding isochrones do not visually change over a large number of iterations. To address this, we limit the number of iterations in each round to 50.
\end{enumerate}

\begin{figure*}
    \centering
    % \sidecaption
\includegraphics[width=1.0\textwidth]{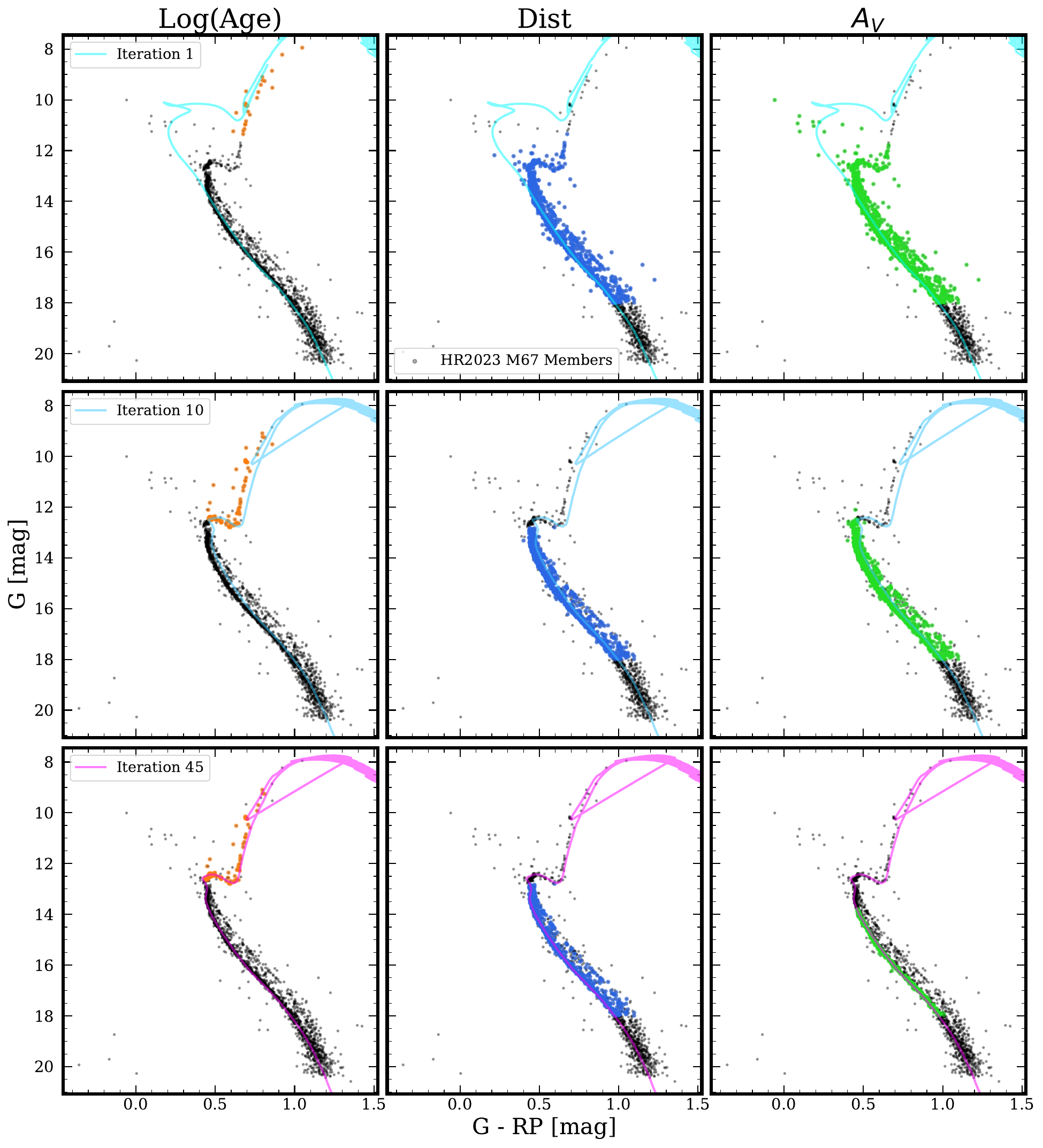}
    \caption{Selection of different subsets of stars based on their evolutionary phases for the estimation of M67’s log(Age) (orange stars), distance (blue stars) and $A_V$ (green stars), at three different iteration steps of the first loop (see Sect.~\ref{subsec:iter_cluster_param}).}
\label{fig:M67_iterative_fitting_3x3}
\end{figure*}

We show this automated selection of different stars based on their evolutionary phases for estimation of the cluster parameters at the 1st, 10th, and 45th (i.e. convergence for the first round) in Fig.~\ref{fig:M67_iterative_fitting_3x3} for M67 (see also Fig.~\ref{fig:cluster_param_iter_method}). \referee{Furthermore, we test the feasibility of this method on simulated clusters in Fig.~\ref{fig:accuracy_of_cluster_params}.}

\section{Comparison with conditional sampling of model parameters posteriors with fixed cluster parameters}\label{app:conditional_sbi}

\begin{figure}
    \centering
    \includegraphics[width=0.5\textwidth]{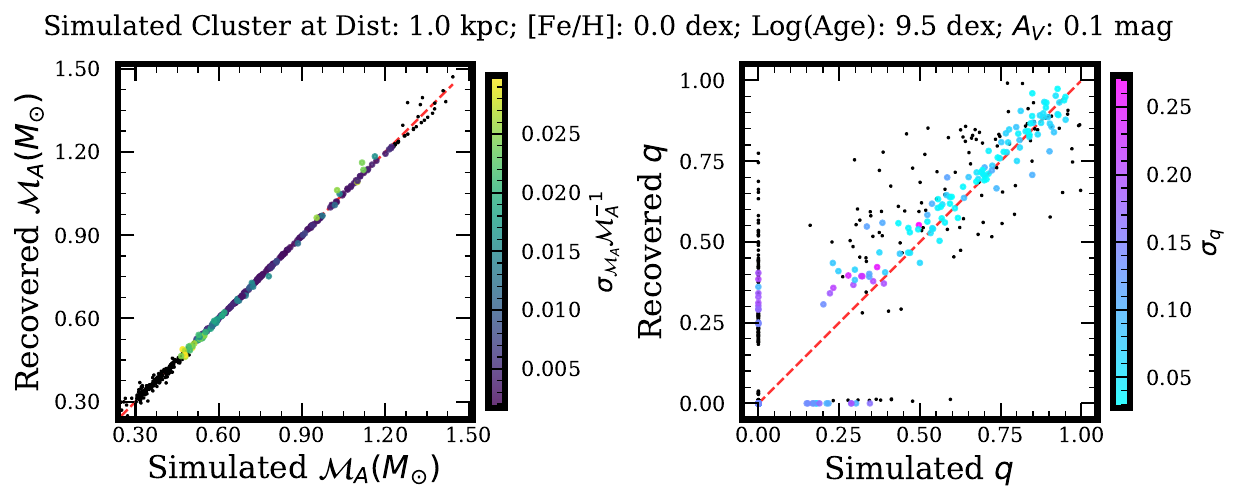}
    \caption{Figure~\ref{fig:clus_performance_sbi} but using conditional sampling of the mass-$q$ posteriors while keeping the cluster parameters fixed.}
    \label{fig:conditional_mass_q_sim_clus_comp}
\end{figure}

The SBI package also comes along with a method to sample conditional distributions, i.e. one can provide conditions on a subset of model parameters while sampling the posterior distributions of the rest. However, as previously mentioned in Sect.~\ref{subsec:apply_cluster_priors}, one of the main disadvantages of this method is that it is computationally very expensive along with the limitation that one can only provide delta functions as conditions, thereby rendering this method worthwhile only when the cluster parameters are known a priori with a very high degree of accuracy. Although we see some improvement in the overall accuracy of the masses and mass-ratios estimates in Fig.~\ref{fig:conditional_mass_q_sim_clus_comp}, we have noted in Fig.~\ref{fig:clus_performance_sbi} that our method can produce satisfactory results at a much lower computational cost without a priori knowledge about the cluster parameters.

\section{Performance for simulated clusters: additional analysis}\label{app:perform_sim_clusters}

\subsection{Performance of iterative cluster parameter estimation method}\label{app:perform_iterative_cluster_param_method}

\begin{figure}[h]
    \centering
    % \sidecaption
\includegraphics[width=0.5\textwidth]{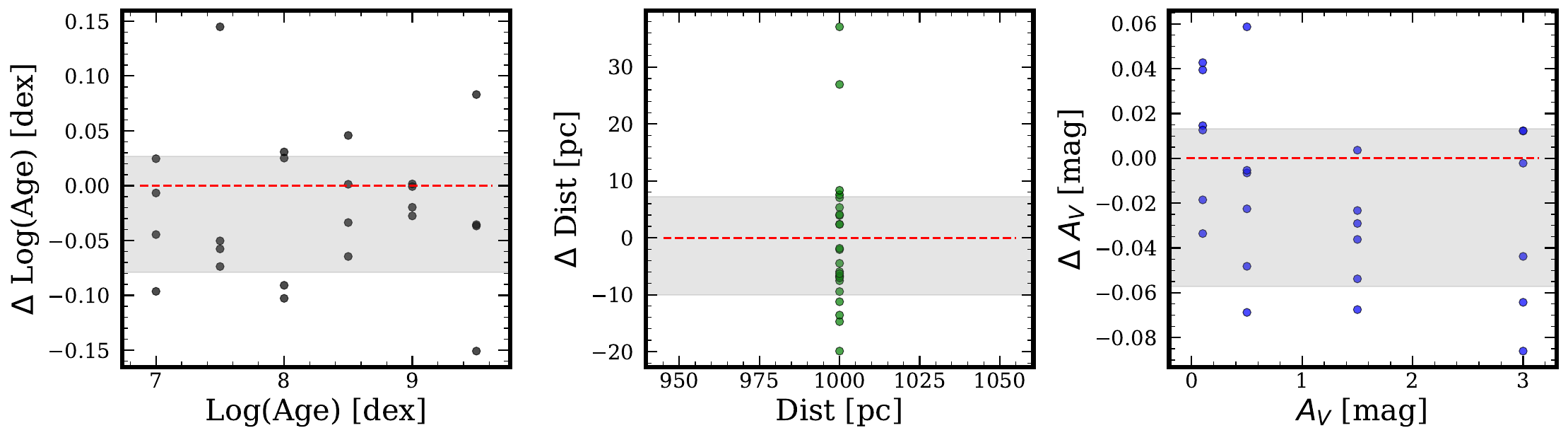}
    \caption{Recovery of cluster parameters for 24 simulated clusters of solar metallicity placed at 1 kpc. The region between the 16th and the 84th percentiles is enclosed by the gray shaded area.}
\label{fig:accuracy_of_cluster_params}
\end{figure}

In this section, we test the viability of the iterative cluster parameter estimation method described in Sect.~\ref{subsec:iter_cluster_param} for simulated clusters. As mentioned earlier, we kept the metallicity [Fe/H] constant, in this case the true value of the simulated cluster, while iterating through other parameters. We show the comparison of the recovered with the simulated parameters for 24 simulated solar metallicity clusters at 1 kpc in Fig.~\ref{fig:accuracy_of_cluster_params} wherein one can observe that the method works well for a wide range of cluster parameters. The predicted log(Age) is within $\pm$ 0.05-0.1 dex of the simulated age of the clusters. On the other hand, the predicted distance is mostly accurate within $\pm$ 20 pc (2\%), whereas the line-of-sight extinction $A_{V}$ is predicted with an accuracy of about 0.08~mag. These are similar to the typical offsets reported in Sect.~\ref{subsec:comp_clus_param_with_lit} compared to the cluster parameters provided in the literature.

 \subsection{Detection efficiency and false positive rate of unresolved binaries above $q_{\mathrm{thresh}}$}\label{app:tpr_fpr_above_q_thresh}

\begin{figure}
    \centering
    \includegraphics[width=0.5\textwidth]{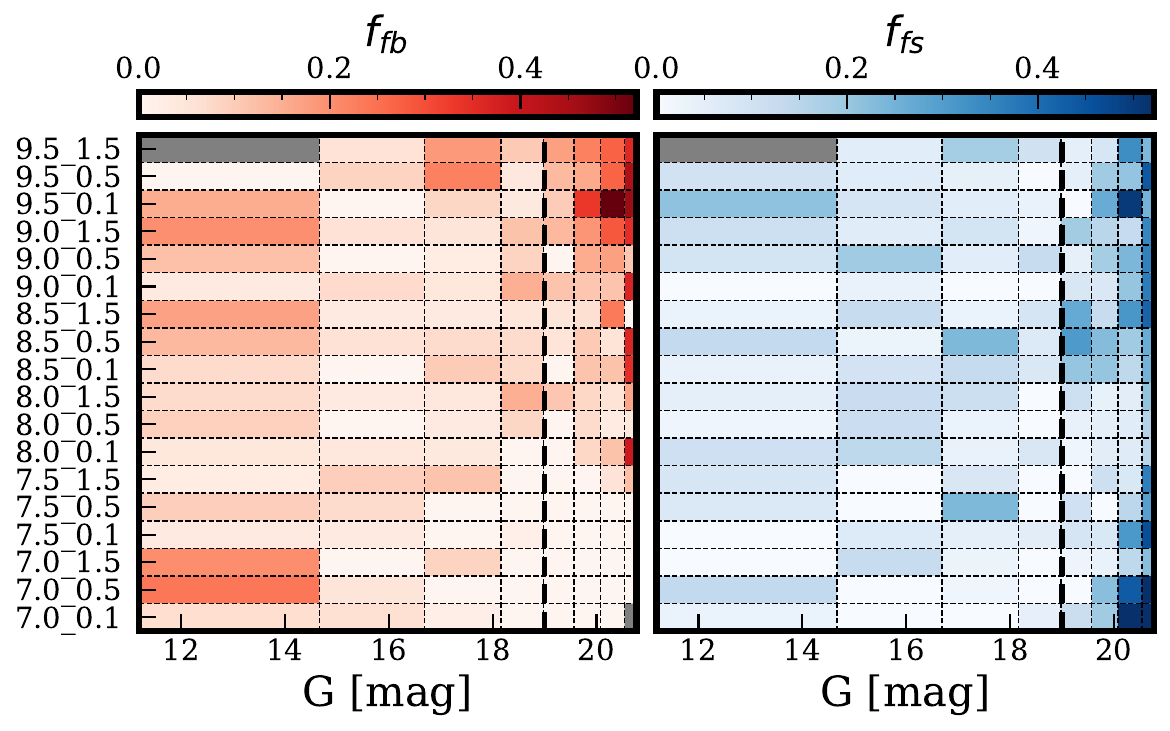}
    \caption{\textit{Left}: Fraction of falsely predicted binaries defined in Eq.~\ref{eq:false_binaries_singles_fraction} in each $G$ mag bin for 24 simulated solar metallicity clusters placed at 1 kpc. The clusters are labelled as "log(Age)\_$A_{V}$". \textit{Right}: Similar to \textit{left} but for fraction of falsely predicted singles. Gray coloured bins indicate zero predicted or true binaries. The bold vertical dashed line indicates $G = 19$ mag.}
\label{fig:false_singles_binaries}
\end{figure}

In Fig.~\ref{fig:clus_performance_sbi}, we observed that our method can lead to a few stars being misclassified as binaries or singles. To see if this false classification is the case for specific stars in the CMD, we consider different bins in the $G$ magnitude, each containing almost similar number of stars. In each bin, we compute the fraction of stars that are falsely predicted as binaries or singles, that is, $f_{fb}$ and $f_{fs}$, respectively, using Eq.~\ref{eq:false_binaries_singles_fraction}.

\begin{equation}
    f_{fb} = \frac{\# \text{false binaries}}{\# \text{predicted binaries}};\quad
    f_{fs} = \frac{\# \text{false singles}}{\# \text{true binaries}}
    \label{eq:false_binaries_singles_fraction}
\end{equation}

where predicted binaries are stars with $q_{\mathrm{pred}} \geq ~\max\left(q_{\mathrm{thresh}}, \frac{0.1}{M_{\mathrm{A}}}\right)$ while false binaries are the subset of predicted binaries with $q_{\mathrm{true}} < q_{\mathrm{thresh}}$ and similarly for false singles and true binaries. Figure~\ref{fig:false_singles_binaries} shows these fractions for the case of 24 simulated solar metallicity clusters placed at 1~kpc, named as "log(Age)\_$A_{V}$", in each $G$ magnitude bin. The gray coloured bins indicate where there were no predicted or true binaries. As one can see, $f_{fb}$ and $f_{fs}$ are mostly lower than 0.2 with a higher false classification rate for stars near the MSTO and those with $G$ mag fainter than 19 mag. This is expected because: a) the single and binary sequences become nearly indistinguishable near the MSTO, and b) there are high uncertainties in the observables for very faint stars, especially the parallax, which becomes significantly uncertain for $G$ $\gtrsim$ 18~mag \citep{GaiaCollaboration2023}.

\subsection{q residuals and uncertainties}\label{app:q_residual_abs_g_mag_true_q}

\begin{figure}
    \centering
    \includegraphics[width=0.5\textwidth]{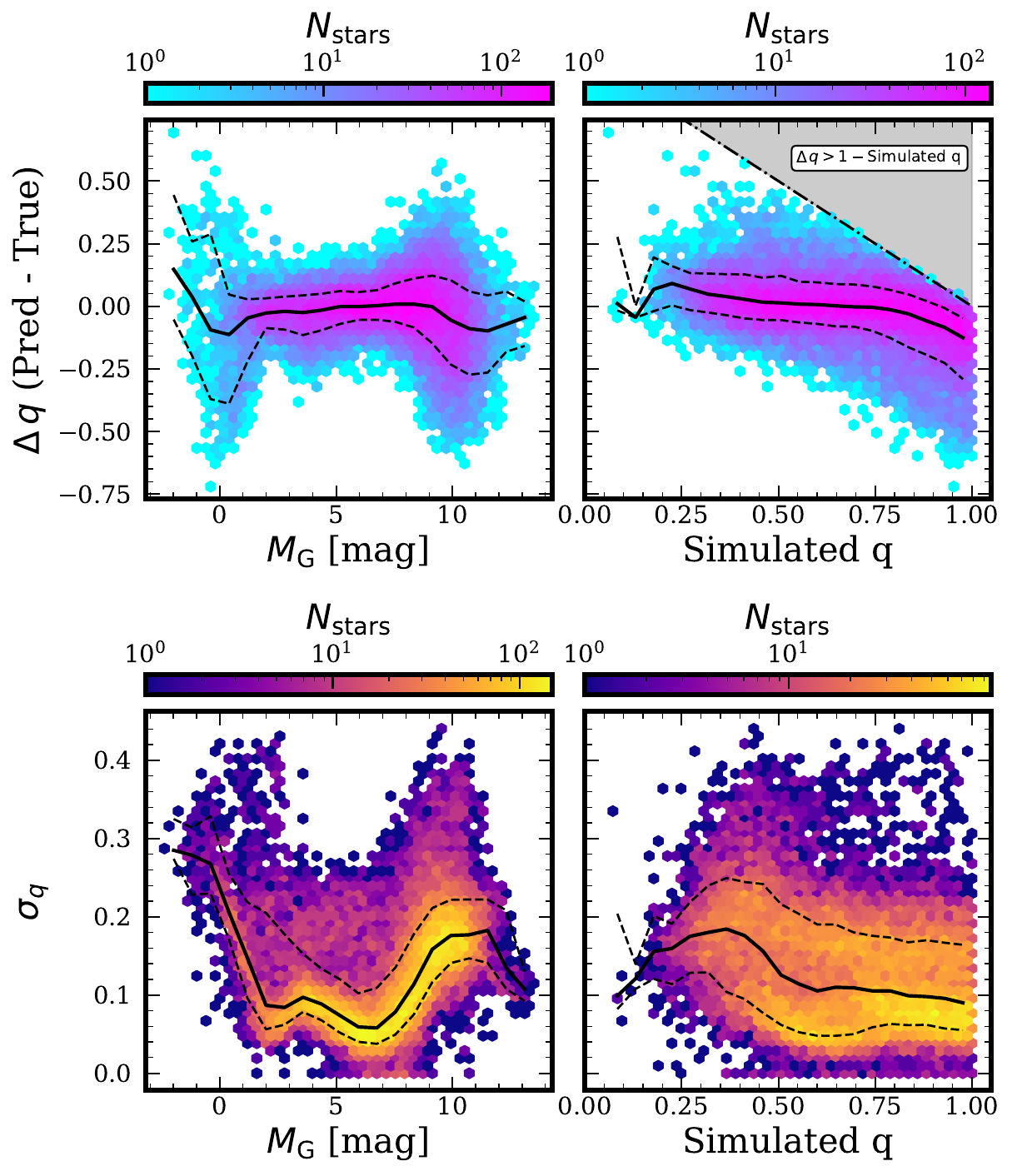}
    \caption{\referee{Dependence of the $q$ residuals (\textit{top}) and predicted uncertainties (\textit{bottom}) on the absolute $G$ magnitude (\textit{left}) and simulated $q$ (\textit{right}). The solid and dashed lines denote the median and 16th–84th percentile ranges, respectively. The shaded region indicates $q_\mathrm{pred} > 1$, values that are unphysical for a binary and therefore not permitted within our framework.}}
    \label{fig:q_residual_vs_abs_g_true_q}
\end{figure}

\referee{We assess the recovery of binary mass ratios for correctly identified unresolved binaries in simulated clusters described in App.~\ref{app:reliably_predict_underlying_bf_vs_mass_A}. Figure~\ref{fig:q_residual_vs_abs_g_true_q} shows the residuals in $q$ (defined as recovered minus simulated) and the predicted uncertainties as functions of absolute $G$ magnitude and simulated mass ratio. The solid and dashed lines represent the median and percentile trends, respectively. The median residuals remain negligible across a broad range of magnitudes and mass ratios but increase for both bright and faint MS stars. Moreover, systems with very high simulated mass ratios tend to show systematically underestimated recovered values. This arises because using the mode of a $q$ posterior truncated at $q=1.0$ as the summary statistic inherently biases the estimates toward lower values. A similar underestimation is seen for very low-mass stars (upper left), which in our simulations preferentially exhibit high $q$ values due to the imposed lower-mass limit of $0.1\,M_\odot$. Moreover, an additional bias may also arise from the $M_\mathrm{A}-q$ degeneracy at high mass-ratios, which remains unaddressed in the absence of a mass-function prior (see Sect.~\ref{sec:results}).}

\referee{Larger residuals for bright MS stars that are closer to the MSTO are also associated with higher predicted uncertainties. This behaviour is expected (and reassuring), as it reflects the intrinsic limitation of the method—near the MSTO, the single and binary-star sequences become nearly indistinguishable, naturally leading to higher uncertainties (see also Fig.~\ref{fig:false_singles_binaries}). Furthermore, broad or multimodal posteriors for a minority of MSTO stars are an expected consequence of the photometric degeneracies, not a prior-driven artefact.}

\subsection{Required accuracy on the cluster parameters to obtain reliable mass-ratios}\label{app:req_cluster_param_accuracy}

\begin{figure}
    \centering
    \includegraphics[width=0.5\textwidth]{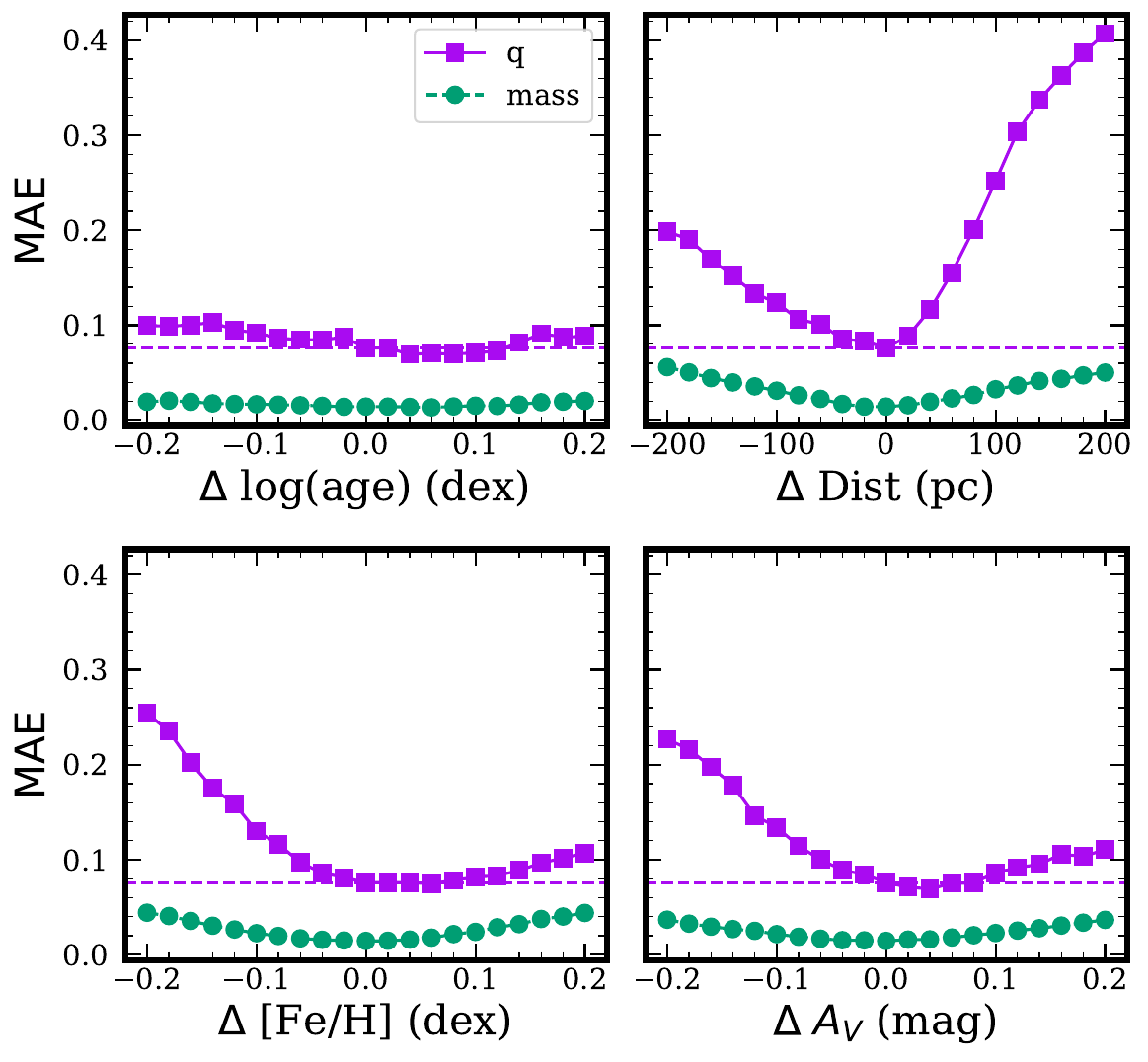}
    \caption{Mean Absolute Error (MAE) in stellar masses and mass-ratios for varying offsets in the mean cluster priors from the true values in case of a solar metallicity cluster simulated at 1~kpc with age 3.5~Gyr and 0.1~$A_V$. The dashed line indicates the MAE in $q$ estimates for an ideal scenario of perfect inference of cluster parameters.}
    \label{fig:mae_vs_offset_in_clus_param}
\end{figure}

We test how inaccuracies in the inferred cluster parameters affect the predictions of mass-ratios and stellar masses in Fig.~\ref{fig:mae_vs_offset_in_clus_param} for a 3.5~Gyr solar metallicity simulated cluster at 1~kpc with a low line-of-sight extinction. We use the mean absolute error (MAE) as the metric to evaluate the overall accuracy. As shown, the offsets provided in App.~\ref{app:perform_iterative_cluster_param_method} are well within the limits wherein the MAE of the mass-ratios and the stellar mass estimations do not change significantly had the cluster parameters been estimated perfectly.

\subsection{Can we reliably predict the underlying variation of binary fraction with stellar mass?}\label{app:reliably_predict_underlying_bf_vs_mass_A}

\begin{figure}
    \centering
    \includegraphics[width=0.42\textwidth]{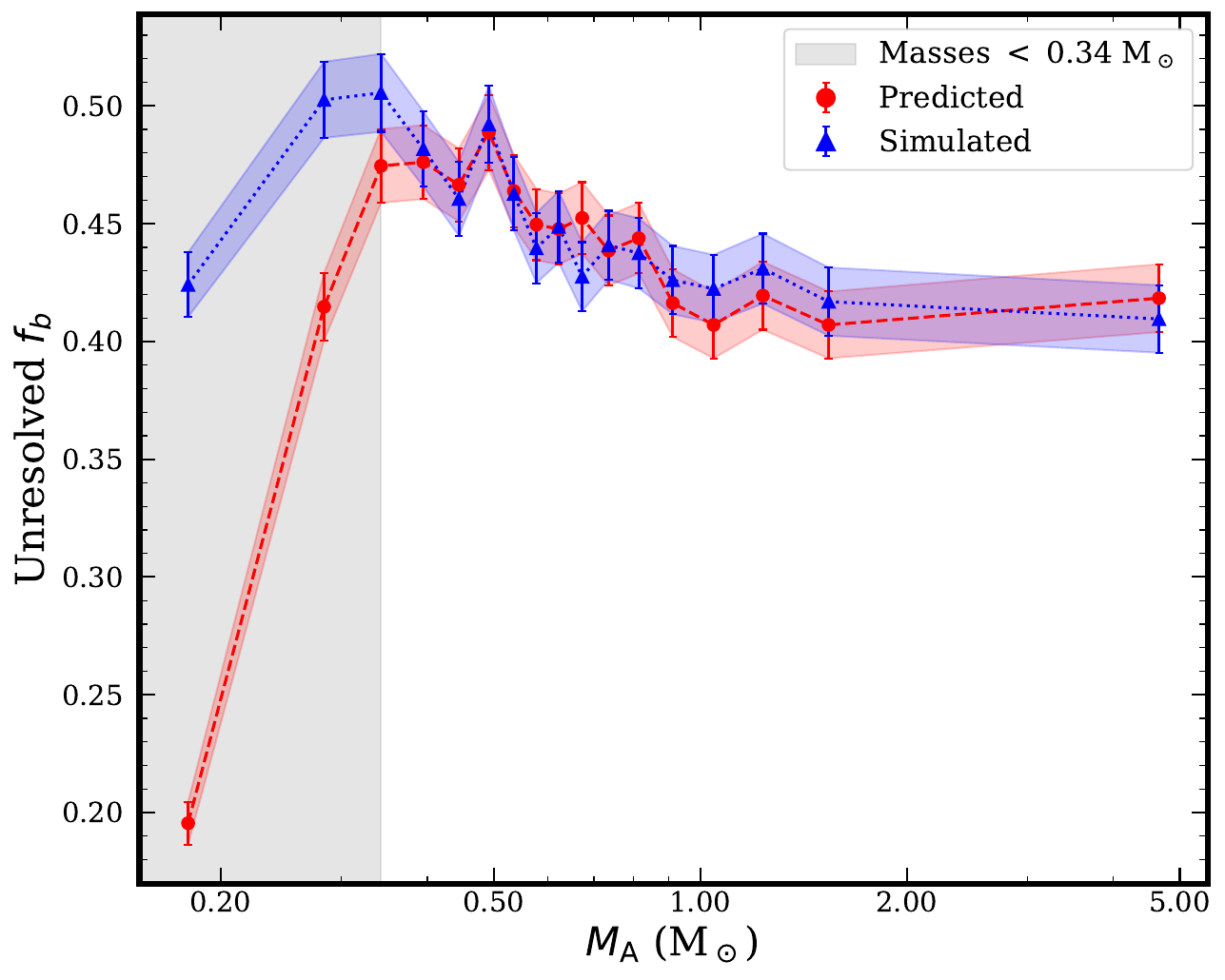}
    \caption{Reliability of the recovered variation of binary fraction w.r.t $M_\mathrm{A}$. We do this test to verify that the observed agreement in Fig.~\ref{fig:binary_vs_mass_stacked} is not dominated by our mass-ratio prior for masses $\gtrsim0.6\,\mathrm{M}_\odot$.}
    \label{fig:test_bf_vs_mass}
\end{figure}

Figure~\ref{fig:binary_vs_mass_stacked} shows good agreement of the recovered binary fractions for different stellar masses with the variation observed in the Galactic field population. However, it remains to be verified that this is not due to the mass-ratio prior, which was motivated by a similar variation (Sect.~\ref{subsec:mass_ratio_prior}), dominating the results. To do such a test, we simulate 42 clusters with similar cluster parameters as the selected clusters in our sample with 2000 members, each with no dependence of the binary fraction and mass-ratio on the stellar mass, while computing the optimal $q_\mathrm{thresh}$ for each cluster. In Fig.~\ref{fig:test_bf_vs_mass}, we show that we can reliably predict the underlying binary fraction versus mass variation for masses $\gtrsim$ 0.34 $\mathrm{M}_\odot$, with differences seen only for masses lower than the mass range ($\lesssim~0.6\,\mathrm{M}_\odot$) which is known to depict discrepancies in the PARSEC isochrones compared to the observed photometry. Hence, our method is capable of recovering the underlying variation in the binary fraction with stellar mass, even when the simulated distribution is independent of mass, demonstrating that the mass-ratio prior does not influence the results in the relevant mass range of $\gtrsim0.6\,\mathrm{M}_\odot$.

\section{Additional figures}

\begin{figure}
    \centering
    \includegraphics[width=0.45\textwidth]{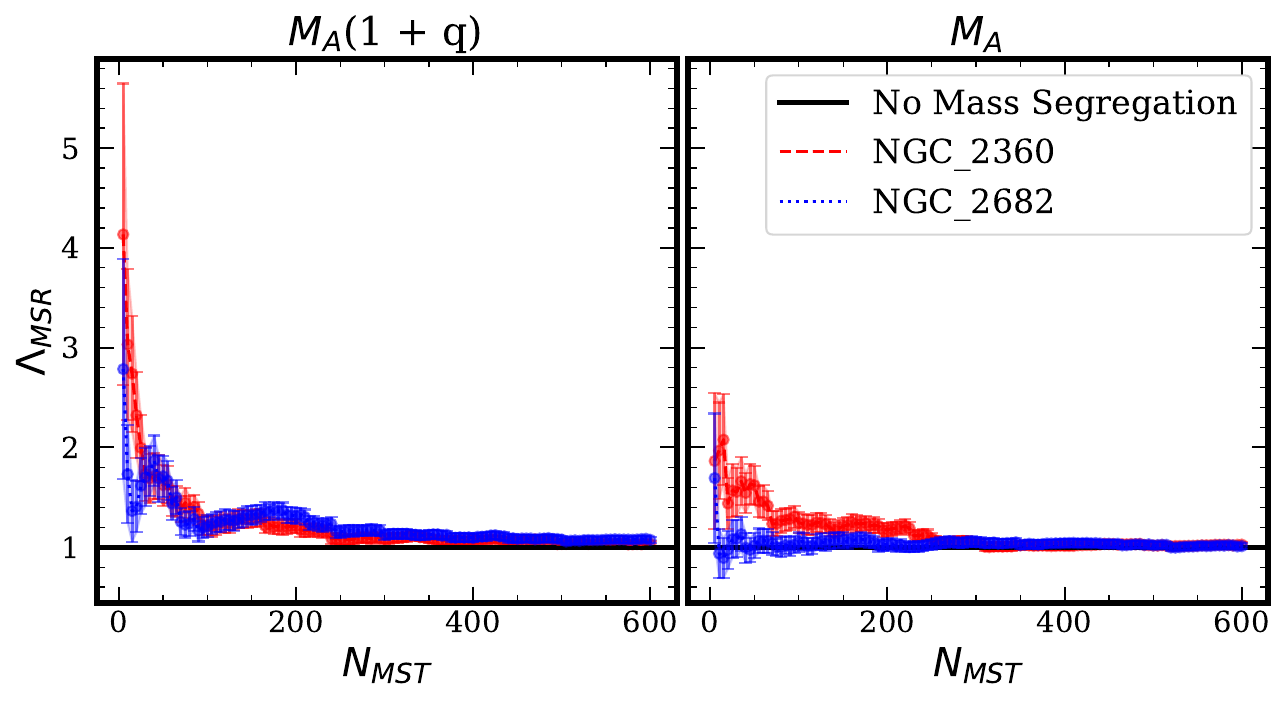}
    \caption{Effect of mass-segregation shown by mass-segregation ratio ($\Lambda_{MSR}$) as a function of the number of stars used to construct the MST ($N_{MST}$). \textit{Left}: The total mass of the binary system is considered while constructing the MST. \textit{Right}: Only the primary mass is used. The solid black line indicates the case of no mass segregation.}
    \label{fig:mass_seg_dual_plot}
\end{figure}

In Fig.~\ref{fig:mass_seg_dual_plot}, we study the signature of mass segregation in NGC 2360 and NGC 2682 and how the inclusion of masses of binary companions affects our conclusion about the mass segregation effect in OCs. We adopt the methodology described by \citet{Allison2009}, constructing a Minimum Spanning Tree (MST) \citep[e.g.][]{Kruskal1956, Prim1957} to quantify the degree of mass segregation. Comparing the length of the MST of the most massive stars in a cluster to the one corresponding to a random spatial distribution, we can compute their ratio as $\Lambda_{MSR}$ for different numbers of stars used to construct the tree, $N_{MST}$. We refer the reader to \citet{Allison2009} for further details about the method. In Fig.~\ref{fig:mass_seg_dual_plot}, we use the total mass of the binary system, $M_{\mathrm{A}} (1+q)$, and only the primary mass, $\mathrm{M}_A$, to construct the MST in the left and right panels of the figure, respectively. We find that accounting for the companion mass significantly alters the inferred degree of mass segregation, particularly in the case of M67. While considering the total mass of unresolved binaries, the cluster is observed to be mass segregated up to approximately the 350 most massive systems. In contrast, using only the primary mass results in the conclusion of a weak mass segregation signature in M67 among only the $\sim20$ most massive stars.

\begin{figure}
    \centering
    \includegraphics[width=0.42\textwidth]{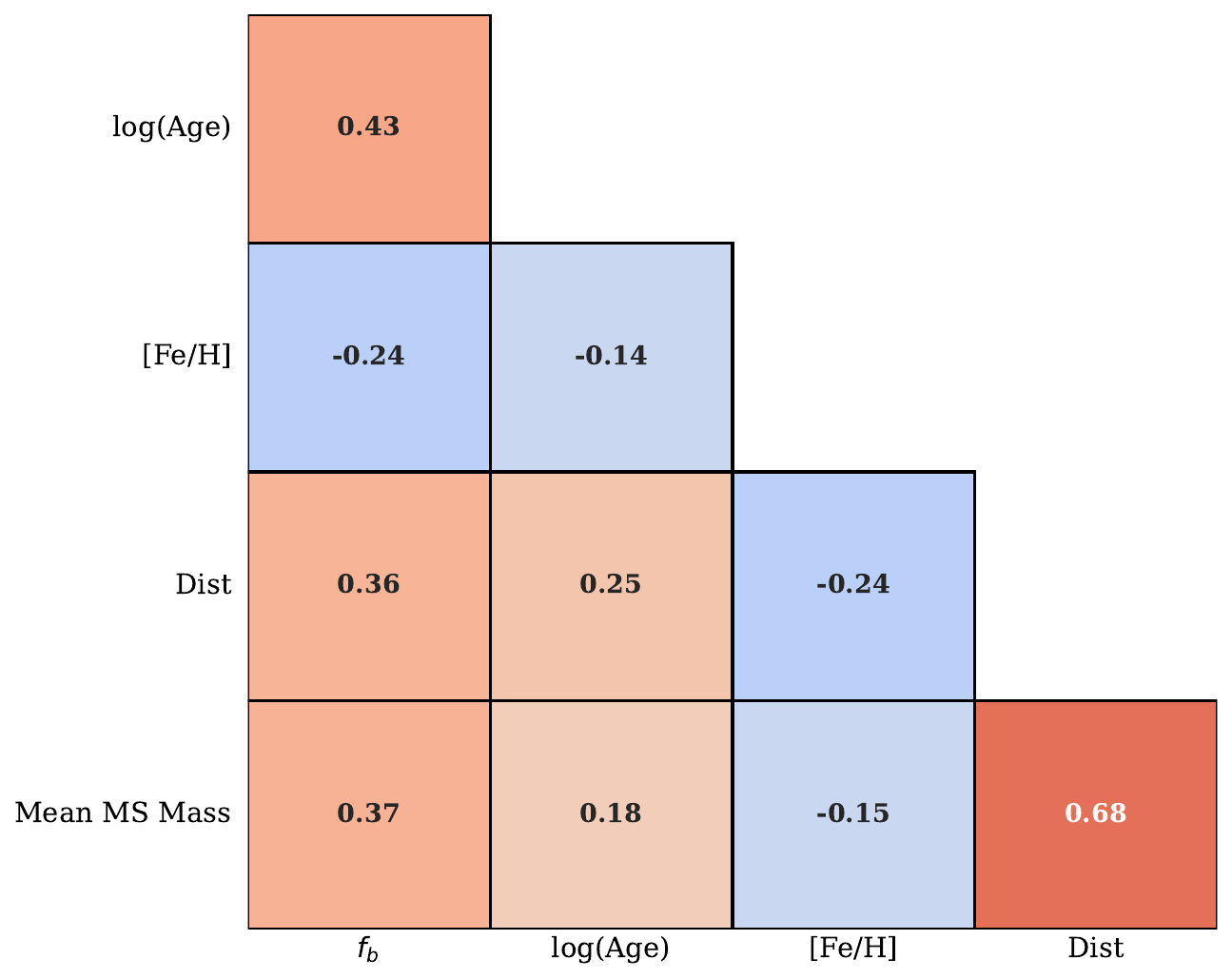}
    \caption{Kendall correlation matrix for the estimated $f_{b}$ ($q\geq0.6$), the cluster parameters viz., log(Age), distance and [Fe/H] along with the mean mass of the MS stars. Also see Fig.~\ref{fig:bf_trends}.}
    \label{fig:bf_corr_matrix}
\end{figure}

\begin{figure}
    \centering
    \includegraphics[width=0.4\textwidth]{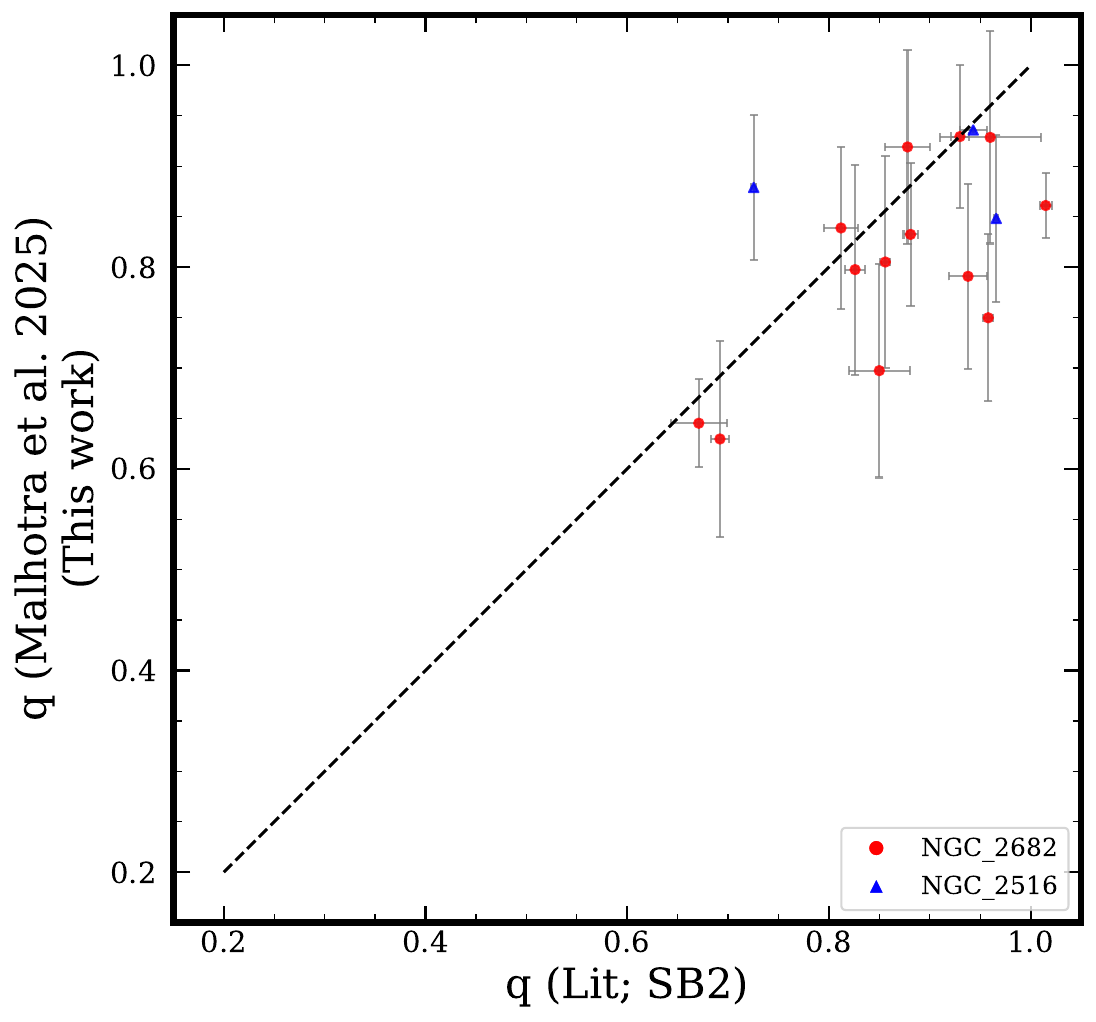}
    \caption{Comparison of the recovered mass-ratios with the corresponding estimates for double-lined spectroscopic binaries (SB2) in M67 (red circles) and NGC 2516 (blue triangles) provided by \citet{Geller2021} and \citet{Lipartito2021} respectively.}
    \label{fig:q_comp_wocs}
\end{figure}

\begin{figure*}
\centering
    \begin{tikzpicture}
       \node[anchor=south west,inner sep=0] (image) at (0,0) {\includegraphics[width=0.99\textwidth]{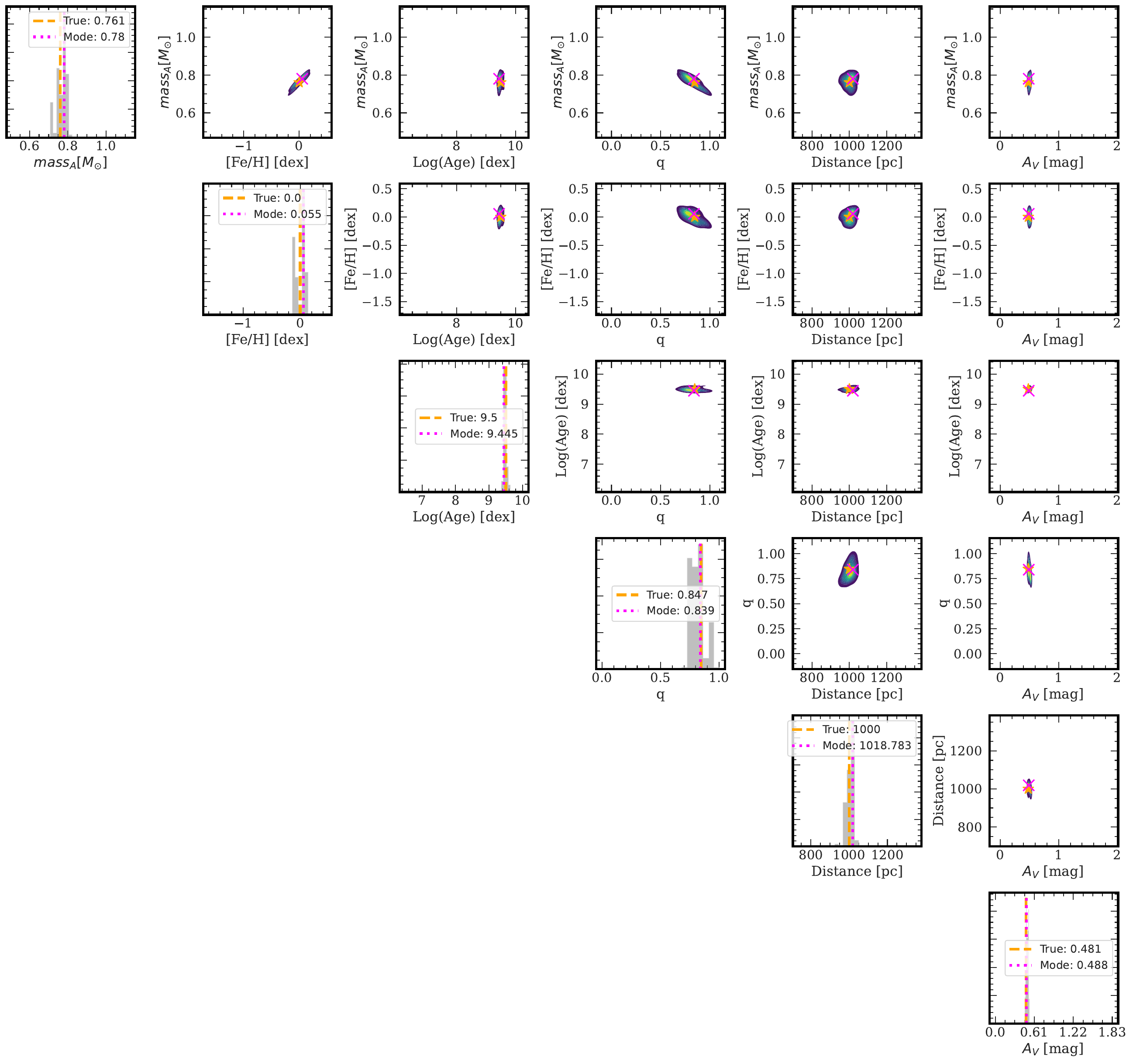}};
       \begin{scope}[x={(image.south east)},y={(image.north west)}]
       \node[anchor=south west,inner sep=0] (image) at (0.0,0.0) {\includegraphics[width=0.5\textwidth]{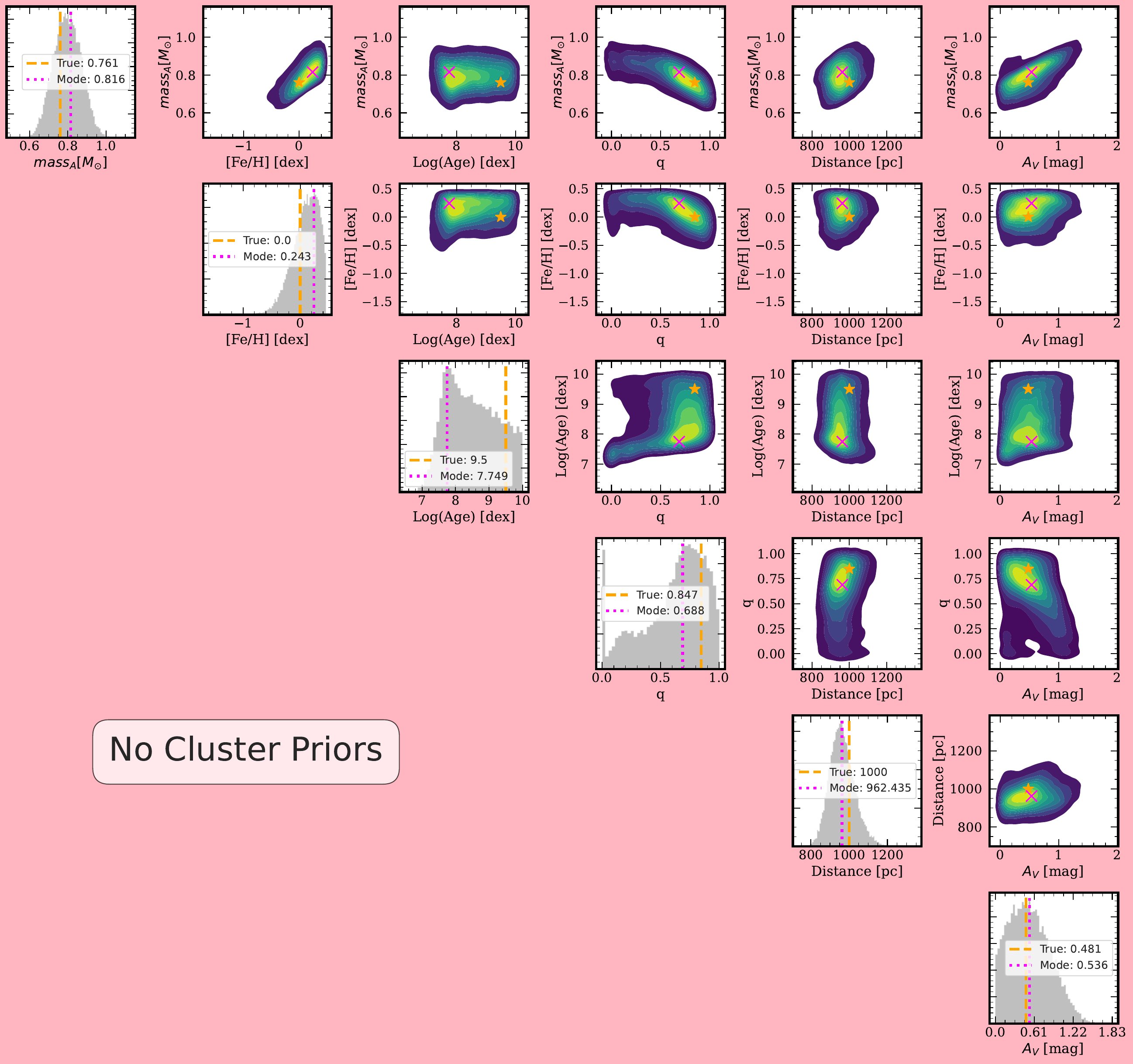}};
       \end{scope}
    \end{tikzpicture}
\caption{Complete version of Fig.~\ref{fig:sim_star_corner_plot} where the unweighted posteriors corresponding to the case of not using the cluster parameter information are shown in the lower shaded plot. The weighted modes estimated for each astrophysical parameter are indicated by the pink dotted lines.} 
\label{fig:sim_star_corner_plot_full} 
\end{figure*}

\begin{figure*}
    \centering
    \includegraphics[width=0.9\textwidth]{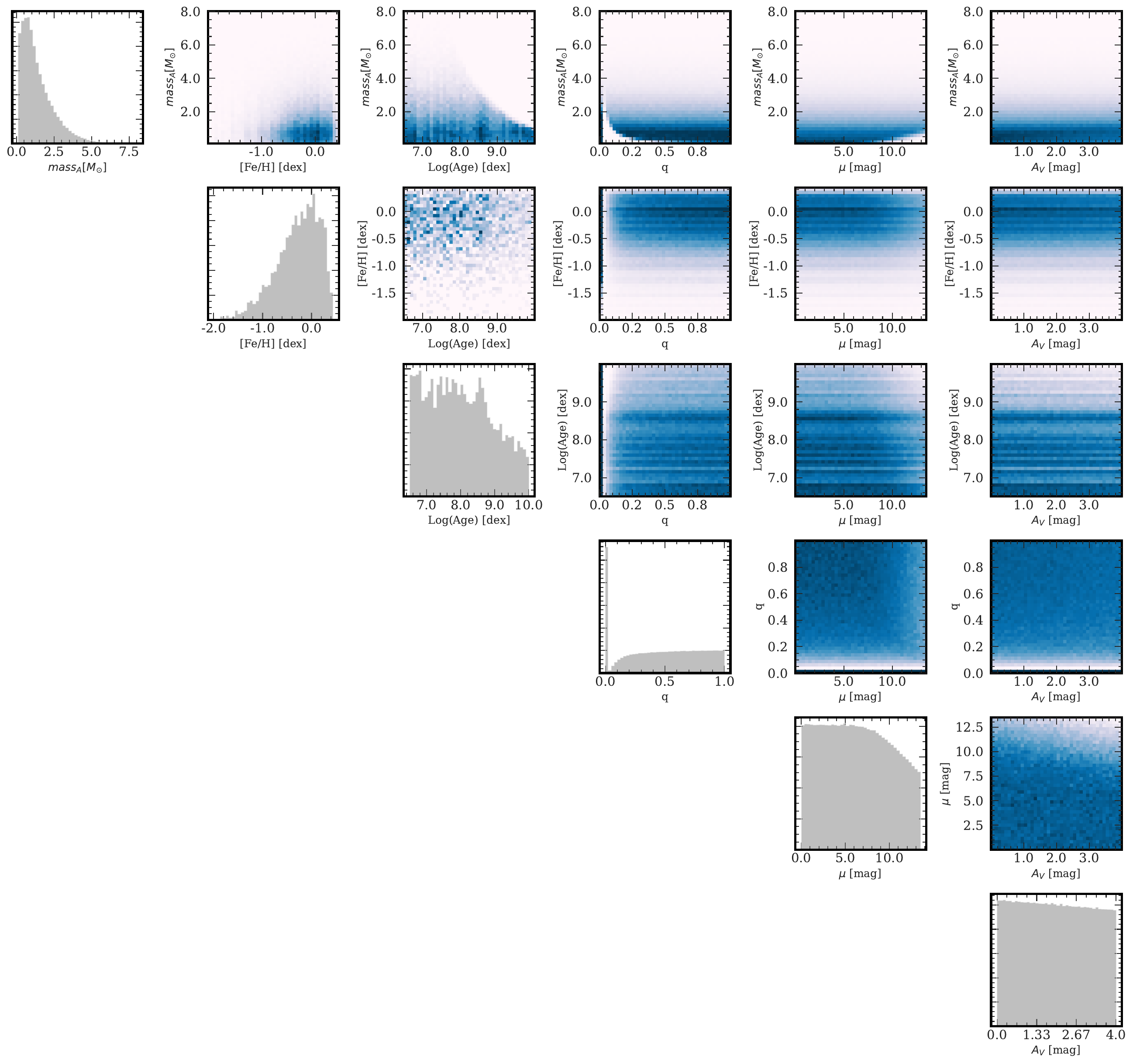}
    \caption{Distribution of model parameters of 2 million simulated single stars and unresolved binary systems.}
    \label{fig:prior_distribution}
\end{figure*}

\section{Published tables}

\begin{table*}[htbp]
\centering
\begin{tabular*}{\textwidth}{@{\extracolsep{\fill}}lccccccccc}
\hline\hline
Cluster & log(Age) & Distance & [Fe/H] & $A_{V}$ & $q_\mathrm{Gaia+}$ & $q_\mathrm{Gaia}$ & $N_\mathrm{binaries}$ & $N_\mathrm{MS}$ & Mass range [$\mathrm{M}_{\odot}$] \\
\hline
ASCC 20 & 7.406 & 361.261 & -0.030 & 0.032 & 0.200 & 0.411 & 66 (50) & 185 & 0.17 - 5.6 \\
Alessi 24 & 8.050 & 483.285 & 0.079 & 0.214 & 0.200 & 0.399 & 120 (95) & 245 & 0.26 - 3.77 \\
BH 99 & 7.644 & 444.880 & -0.099 & 0.305 & 0.207 & 0.385 & 266 (208) & 549 & 0.19 - 5.16 \\
Blanco 1 & 8.445 & 237.525 & 0.040 & 0.019 & 0.200 & 0.393 & 163 (142) & 694 & 0.18 - 3.01 \\
COIN-Gaia 12 & 8.779 & 962.664 & -0.033 & 0.370 & 0.272 & 0.543 & 74 (33) & 120 & 0.47 - 2.08 \\
COIN-Gaia 24 & 8.426 & 978.266 & 0.011 & 0.874 & 0.309 & 0.515 & 52 (34) & 86 & 0.57 - 2.57 \\
Collinder 135 & 7.678 & 294.689 & -0.011 & 0.045 & 0.200 & 0.378 & 86 (63) & 195 & 0.17 - 3.6 \\
Melotte 111 & 8.794 & 86.227 & -0.050 & 0.025 & 0.200 & 0.415 & 84 (64) & 251 & 0.12 - 2.17 \\
NGC 1039 & 8.400 & 500.549 & 0.070 & 0.195 & 0.282 & 0.456 & 388 (287) & 891 & 0.3 - 3.04 \\
NGC 1901 & 8.987 & 418.845 & -0.145 & 0.173 & 0.273 & 0.461 & 93 (81) & 198 & 0.25 - 1.81 \\
NGC 2215 & 8.809 & 958.659 & -0.087 & 0.735 & 0.246 & 0.443 & 112 (92) & 173 & 0.47 - 2.07 \\
NGC 2287 & 8.313 & 735.823 & -0.130 & 0.203 & 0.275 & 0.411 & 481 (405) & 836 & 0.36 - 3.23 \\
NGC 2301 & 8.189 & 886.492 & -0.008 & 0.203 & 0.200 & 0.450 & 495 (342) & 764 & 0.4 - 3.25 \\
NGC 2360 & 9.031 & 1076.271 & -0.030 & 0.306 & 0.340 & 0.548 & 439 (311) & 940 & 0.49 - 1.81 \\
NGC 2447 & 8.806 & 1027.628 & -0.050 & 0.068 & 0.276 & 0.523 & 409 (254) & 875 & 0.47 - 2.12 \\
NGC 2482 & 8.601 & 1360.312 & -0.070 & 0.310 & 0.273 & 0.506 & 99 (71) & 164 & 0.58 - 2.43 \\
NGC 2516 & 8.168 & 411.022 & 0.080 & 0.276 & 0.287 & 0.415 & 1116 (918) & 2574 & 0.25 - 3.55 \\
NGC 2539 & 8.870 & 1316.379 & 0.066 & 0.148 & 0.375 & 0.520 & 246 (177) & 591 & 0.56 - 2.08 \\
NGC 2682 & 9.550 & 864.246 & 0.020 & 0.113 & 0.254 & 0.556 & 510 (345) & 1238 & 0.44 - 1.2 \\
NGC 3114 & 8.158 & 990.941 & 0.050 & 0.243 & 0.200 & 0.506 & 829 (451) & 1444 & 0.45 - 3.74 \\
NGC 3532 & 8.518 & 478.143 & -0.034 & 0.193 & 0.269 & 0.477 & 1563 (1103) & 2627 & 0.27 - 2.75 \\
NGC 3680 & 9.243 & 1049.457 & -0.100 & 0.248 & 0.371 & 0.547 & 48 (39) & 94 & 0.5 - 1.51 \\
NGC 5460 & 8.283 & 708.008 & 0.050 & 0.330 & 0.249 & 0.478 & 127 (82) & 236 & 0.45 - 3.25 \\
NGC 5822 & 8.966 & 827.819 & 0.020 & 0.435 & 0.254 & 0.514 & 423 (254) & 642 & 0.49 - 1.87 \\
NGC 6475 & 8.445 & 277.753 & 0.140 & 0.246 & 0.264 & 0.398 & 642 (515) & 1354 & 0.24 - 2.84 \\
NGC 6633 & 8.932 & 392.501 & -0.043 & 0.287 & 0.261 & 0.441 & 317 (283) & 412 & 0.25 - 2.02 \\
NGC 6811 & 8.991 & 1150.717 & -0.059 & 0.252 & 0.330 & 0.510 & 183 (133) & 356 & 0.49 - 1.85 \\
NGC 6866 & 8.832 & 1444.743 & 0.000 & 0.448 & 0.357 & 0.493 & 207 (172) & 332 & 0.61 - 2.06 \\
NGC 6940 & 8.989 & 1009.287 & 0.009 & 0.445 & 0.277 & 0.530 & 464 (340) & 678 & 0.5 - 1.83 \\
NGC 6991 & 9.183 & 566.730 & 0.003 & 0.273 & 0.317 & 0.467 & 192 (163) & 298 & 0.36 - 1.61 \\
NGC 7058 & 7.803 & 365.789 & 0.010 & 0.165 & 0.200 & 0.416 & 137 (109) & 348 & 0.21 - 4.47 \\
NGC 7063 & 7.873 & 669.290 & -0.090 & 0.346 & 0.220 & 0.419 & 78 (51) & 161 & 0.34 - 4.12 \\
NGC 752 & 9.244 & 439.039 & -0.082 & 0.141 & 0.312 & 0.393 & 139 (130) & 340 & 0.26 - 1.51 \\
RSG 5 & 7.804 & 334.673 & 0.105 & 0.032 & 0.200 & 0.419 & 68 (43) & 157 & 0.18 - 4.4 \\
Ruprecht 147 & 9.447 & 304.350 & 0.120 & 0.247 & 0.291 & 0.426 & 86 (79) & 194 & 0.29 - 1.29 \\
Stephenson 1 & 7.819 & 356.353 & 0.201 & 0.035 & 0.200 & 0.437 & 128 (92) & 294 & 0.22 - 3.74 \\
Stock 1 & 8.669 & 405.869 & 0.120 & 0.341 & 0.200 & 0.454 & 152 (110) & 239 & 0.3 - 2.27 \\
UPK 136 & 7.938 & 640.145 & -0.047 & 0.270 & 0.200 & 0.453 & 55 (39) & 86 & 0.37 - 4.2 \\
UPK 303 & 8.498 & 210.186 & 0.118 & 0.024 & 0.200 & 0.430 & 73 (54) & 236 & 0.2 - 2.86 \\
UPK 429 & 8.413 & 878.987 & 0.018 & 0.123 & 0.234 & 0.543 & 44 (17) & 92 & 0.5 - 2.0 \\
UPK 467 & 8.111 & 590.406 & 0.163 & 0.048 & 0.200 & 0.409 & 51 (33) & 144 & 0.35 - 3.21 \\
UPK 612 & 8.122 & 223.865 & 0.021 & 0.144 & 0.200 & 0.371 & 60 (41) & 212 & 0.18 - 2.24 \\
\hline
\end{tabular*}
\caption{Inferred cluster parameters: log(Age), distance, metallicity [Fe/H] (fixed by spectroscopy; \citealt{Joshi2024}), line-of-sight extinction $A_V$, minimum mass-ratio thresholds $q_\mathrm{Gaia+}$ (when infra-red photometry is available) and $q_\mathrm{Gaia}$ (when only {\it Gaia} DR3 photometry is used), the number of detected unresolved binaries (with those having $q \geq q_\mathrm{Gaia}$ in parentheses), and the number of main-sequence stars with $G \leq 19$~mag in the mass range listed in the last column. The complete table with all columns is available via the CDS.}
\label{tab:cluster_summary}
\end{table*}

\begin{table*}[htbp]
\centering
\begin{tabular*}{\textwidth}{@{\extracolsep{\fill}}lcccccccc}
\hline\hline
Cluster & source\_id & Posterior & $MassA_\mathrm{wtMode}$ [M$_\odot$] & $q_\mathrm{wtMode}$ & UsedInEst & $1 - P_\mathrm{b}$ & $N_\mathrm{samples}$ & $q_\mathrm{thresh}$ \\
\hline
ASCC\_20 & ... & All\_Mags & 5.602 & 0.754 & logage,dist & 0.003 & 3891 & 0.200 \\
ASCC\_20 & ... & Gaia\_TMASS & 4.353 & 0.820 & logage,dist & 0.002 & 3867 & 0.200 \\
... & ... & ... & ... & ... & ... & ... & ... & ... \\
UPK\_612 & ... & Gaia & 1.017 & 0.416 & logage,dist & 0.074 & 599 & 0.371 \\
UPK\_612 & ... & Gaia\_TMASS & 0.202 & 0.000 & logage & 1.000 & 786 & 0.200 \\
\hline
\end{tabular*}
\caption{Preview of the final table containing cluster members' source IDs,            posterior type used, weighted mode of $M_\mathrm{A}$ and $q$,            used in estimating which cluster parameters, probability of being single,            number of valid samples, and mass ratio threshold applied. The full             version containing all stars and all columns is available via the CDS.}
\label{tab:dataset}
\end{table*}

\end{appendix}

%------------------------------------------------------------------
\end{document}